\documentclass{aa}  
\usepackage{txfonts}
\usepackage{graphicx}
\usepackage{gensymb}
\usepackage{multirow, makecell}
\usepackage{pdflscape}
\usepackage{hyperref}
\hypersetup{
    colorlinks=true,
    linkcolor=blue,
    filecolor=blue,      
    urlcolor=blue,
    citecolor=blue,
}

\begin{document}

\title{Onboard catalogue of known X-ray sources for \textit{SVOM}/ECLAIRs}
\subtitle{}

\author{N. Dagoneau\inst{1} \and
        S. Schanne\inst{1} \and
        J. Rodriguez\inst{1} \and 
        J.-L. Atteia\inst{2} \and
        B. Cordier\inst{1}}

\institute{Lab AIM - CEA, CNRS, Université Paris-Saclay, Université de Paris, 91191 Gif-sur-Yvette, France\\
           \email{nicolas.dagoneau@cea.fr}
           \and
            IRAP, Université de Toulouse, CNES, CNRS, UPS, Toulouse, France}

\date{Received XXX; accepted XXX}

\abstract{The \textit{SVOM} mission under development will carry various instruments, and in particular the coded-mask telescope ECLAIRs, with a large field of view of about 2 sr, operating in the 4--150 keV energy band, whose goal is to detect high energy transients such as gamma-ray bursts. The trigger software onboard ECLAIRs will search for new hard X-ray sources appearing in the sky, as well as peculiar behaviour (e.g. strong outbursts) from known sources, in order to repoint the satellite to perform follow-up observations with its onboard narrow field of view instruments. The presence of known X-ray sources must be disentangled from the appearance of new sources. This is done with the help of an onboard source catalogue, which we present in this paper. As an input we use catalogues of X-ray sources detected by \textit{Swift}/BAT and \textit{MAXI}/GSC and we study the influence of the sources on ECLAIRs' background level and on the quality of the sky image reconstruction process. We show that the influence of the sources depends on the pointing direction on the sky, on the energy band and on the exposure time. In the Galactic centre, the known sources contribution largely dominates the cosmic X-ray background, which is, on the contrary, the main background in sky regions empty of strong sources. We also demonstrate the need to clean the sources contributions in order to maintain a low noise level in the sky images and to keep the threshold applied for the detection of new sources as low as possible, without introducing false triggers. We briefly describe one of our cleaning methods and its challenges. Finally, we present the overall structure of the onboard catalogue and the way it will be used to perform the source cleaning and disentangle the detections of new sources from outbursts of known sources.}


\maketitle
\section{Introduction}

\subsection{The \textit{SVOM} mission}

The most violent and energetic phenomena of the Universe usually emit copious amounts of high-energy radiation (typically X- and gamma-ray photons). These events are also transient, unpredictable and some can occur on very short (sub-second) time scales. Dedicated instruments and specific observational strategies are therefore needed to catch those events. \textit{SVOM} (Space-based multi-band astronomical Variable Objects Monitor, \citealt{wei_deep_2016}) is a French-Chinese mission dedicated to the detection and follow-up of short flashes of hard X-ray and gamma-ray photons called gamma-ray bursts (GRB) and of other high-energy transients. \textit{SVOM} is currently under development and planned to be operational after mid-2022.

\begin{figure}
\resizebox{\hsize}{!}{\includegraphics{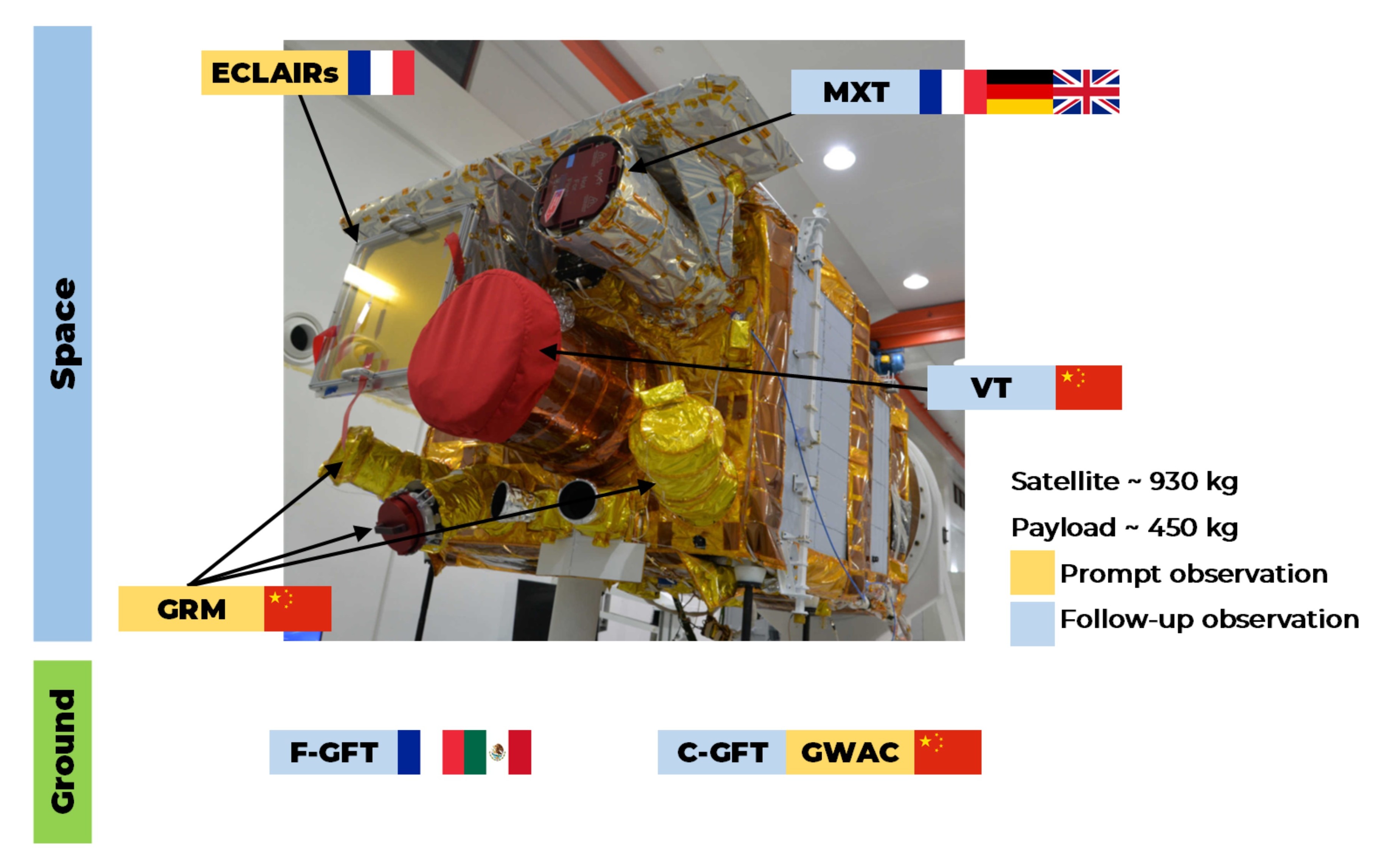}}
\caption{Overview of the \textit{SVOM} platform. The wide field of view instruments are shown in orange boxes. The small field of view instruments are depicted in blue boxes. The photo shows the structural and thermal models of the instruments on the qualification model of the satellite (CNES/SECM).}
\label{fig:svom}
\end{figure}

The \textit{SVOM} project involves both space-based and ground-based telescopes. 
Figure \ref{fig:svom} shows the \textit{SVOM} space platform which includes four space instruments: the ECLAIRs telescope, the Gamma Ray-burst Monitor (GRM), the Microchannel X-ray Telescope (MXT) and the Visible Telescope (VT). 
The \textit{SVOM} mission also includes ground-based telescopes: a set of Ground Wide Angle Cameras (GWAC) and two dedicated Ground Follow-up Telescopes (GFTs), reported schematically on this figure as well as the main contributing countries. 

The goal of the \textit{SVOM} \textit{core program} is the multi-wavelength study of GRBs, including the prompt and the afterglow emission, in order to obtain a complete sample of well-characterised GRBs including redshift measurements for about $2/3$ of the total sample. First, the GRB prompt emission is detected and its gamma-ray light-curve and spectrum are characterised with the wide field instruments ECLAIRs and GRM onboard, and simultaneously studied in the visible band with the GWAC on-ground, which points as much as possible at the same sky region as the onboard wide field instruments ECLAIRs and GRM. After the onboard GRB detection and a first automatic estimate of its best celestial coordinates by ECLAIRs, the spacecraft will slew to the GRB position, and the afterglow will be observed by the narrow field instruments MXT in the X-rays and VT in the visible band. Simultaneously the GRB alert including its position will be transmitted to the ground through a VHF antenna, and the GFTs will perform follow-up observations in the visible and near-infrared bands. The GRB alerts and their subsequent refined positions obtained by the \textit{SVOM} follow-up telescopes are quickly and publicly distributed to the whole community of interested observers of the transient sky. This strategy will enhance the number of accurately localised bursts, which is essential to perform fruitful follow-up observations with large ground-based facilities, including spectrometers on large ground-based telescopes, important to determine the redshift of those events.

Besides the GRB detection and observation as part of its \textit{core program}, \textit{SVOM} will also carry out pre-planned observations in its so called \textit{general program}. This program includes known-source observations with the narrow field of view instruments (MXT, VT and GFTs) such as X-ray binaries, AGNs, blazars, ultra-luminous X-ray sources or cataclysmic variables, as well as wide field surveys with ECLAIRs and GRM. These observations will address some questions about the physics behind the emission processes in these objects (accretion through disks, ejection in jets). Also, following detection by other instruments, \textit{SVOM} will perform fast observations of targets of opportunity for transient follow-up or optical counterpart searches of multi-messenger events. The \textit{SVOM} prospects on GRB science, rapid follow-up observations and observatory science are more detailed in \cite{wei_deep_2016}.

\subsubsection{Orbit and pointing}
\label{sec:orbit_pointing}

The \textit{SVOM} spacecraft will be placed in a quasi-circular low-earth orbit (altitude $\approx 625$ km) with an inclination of about 30$\degree$.

Its attitude (orientation with respect to an inertial reference frame) was optimised to follow the so-called ``B1 law'' which ensures an anti-solar pointing in order to protect the payload from the Sun light and guarantees that the field of view is simultaneously observable by ground instruments in the night hemisphere of the Earth. This law also avoids the presence in ECLAIRs' field of view of the Galactic plane (Galactic latitudes $|b|< 10\degree$) and the very bright X-ray source Scorpius X-1 (Sco X-1 in the following) with a margin of 1$\degree$. Consequently to this orbit and pointing strategy, the Earth will regularly cross the field of view of the instruments and influence their background level and shape. In our study, neither the Sun avoidance constraint nor the margin around Sco X-1 is considered and the following figures and values may differ slightly from those given in other communications. In the nominal phase of the mission (the first three years), the observing time outside the considered B1 law is 60$\%$ of the useful mission time and increases to more than 74$\%$ in the extended phase. 

\begin{figure}
\resizebox{\hsize}{!}{\includegraphics{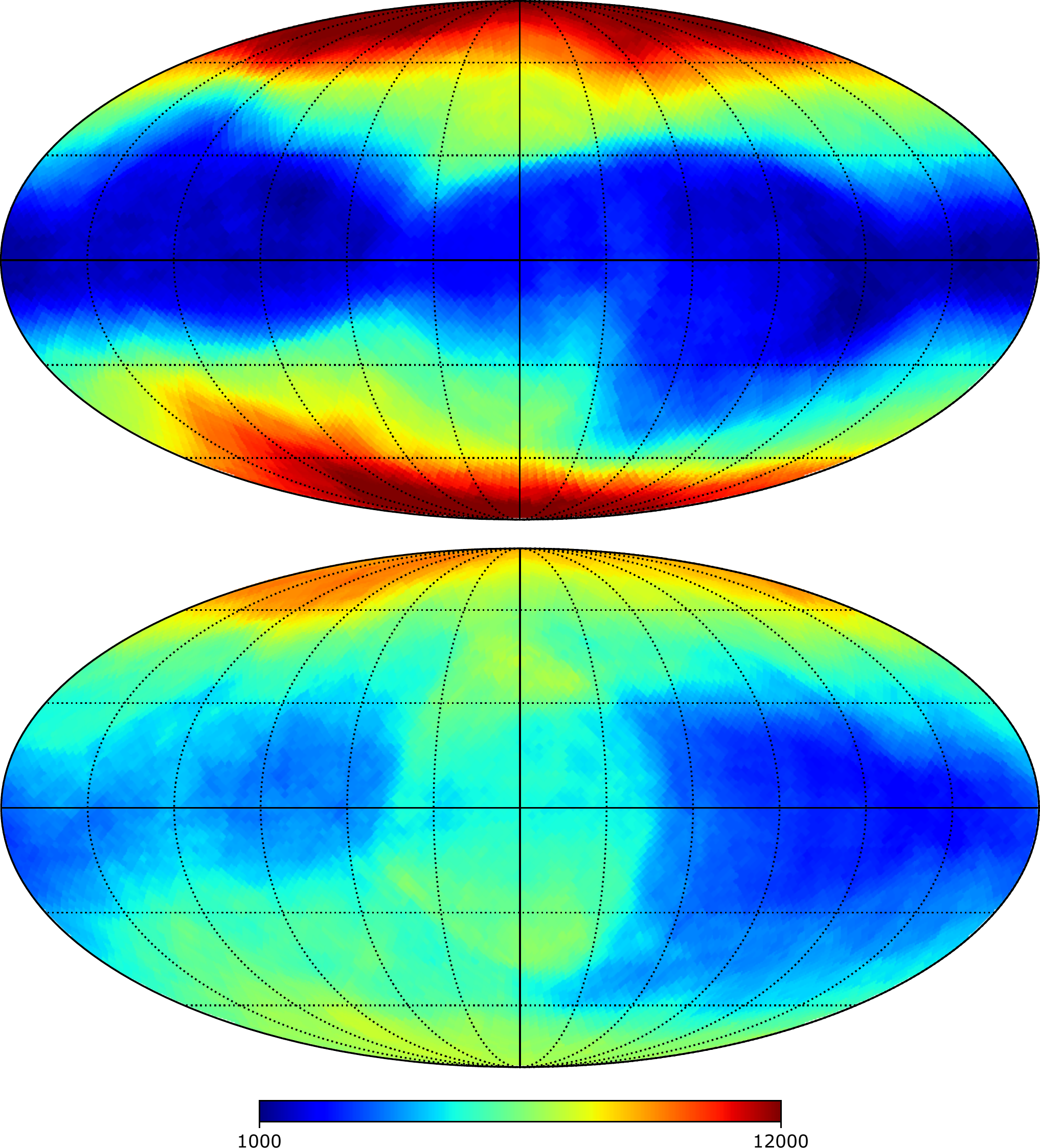}}
\caption{Map of the exposure time in a one year simulation by CNES \citep{jaubert_realistic_2017} in kilo-seconds represented in Galactic coordinates (longitude increasing from right to left). Top: nominal phase, bottom: extended phase.}
\label{fig:exposure}
\end{figure}

\begin{figure}
\resizebox{\hsize}{!}{\includegraphics{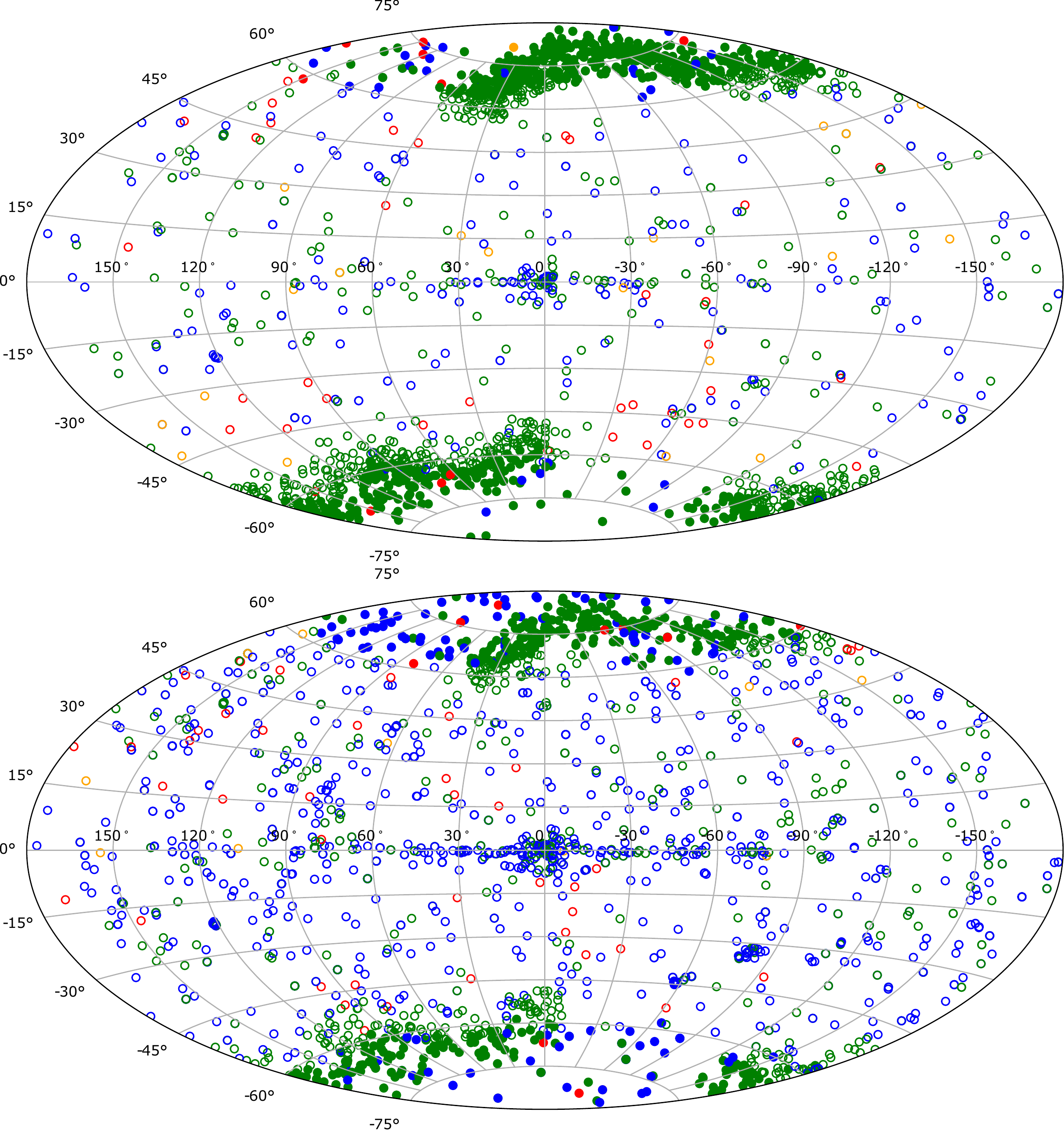}}
\caption{Distribution of the pointing directions on the sky in Galactic coordinates in a one year simulation by CNES \citep{jaubert_realistic_2017}. Top: nominal phase, bottom: extended phase. Filled circles: pointings inside the B1 law, empty circles: pointings outside the B1 law. Green: \textit{general program} pointings, blue: nominal targets of opportunity (transient follow-up, GRB revisits), orange: exceptional targets of opportunity (exceptional astronomical events
requiring fast follow-up), red: GRB follow-up pointings.}
\label{fig:pointings}
\end{figure}

Figure \ref{fig:exposure} represents the expected exposure maps for the nominal (first three years) and extended phases of the mission, while Fig.~\ref{fig:pointings} is the translation in terms of individual pointings (attitude of the instruments optical axis). Even if the Galactic poles have a much larger exposure time during the nominal phase than the Galactic centre (Fig.~\ref{fig:exposure} top), this dichotomy is smoothed during the extended phase (Fig.~\ref{fig:exposure} bottom), and the number of pointings outside the B1 law, in particular toward the Galactic centre and plane, is significantly increased (Fig.~\ref{fig:pointings}).

\subsection{The ECLAIRs telescope}

\begin{figure}
\resizebox{\hsize}{!}{\includegraphics{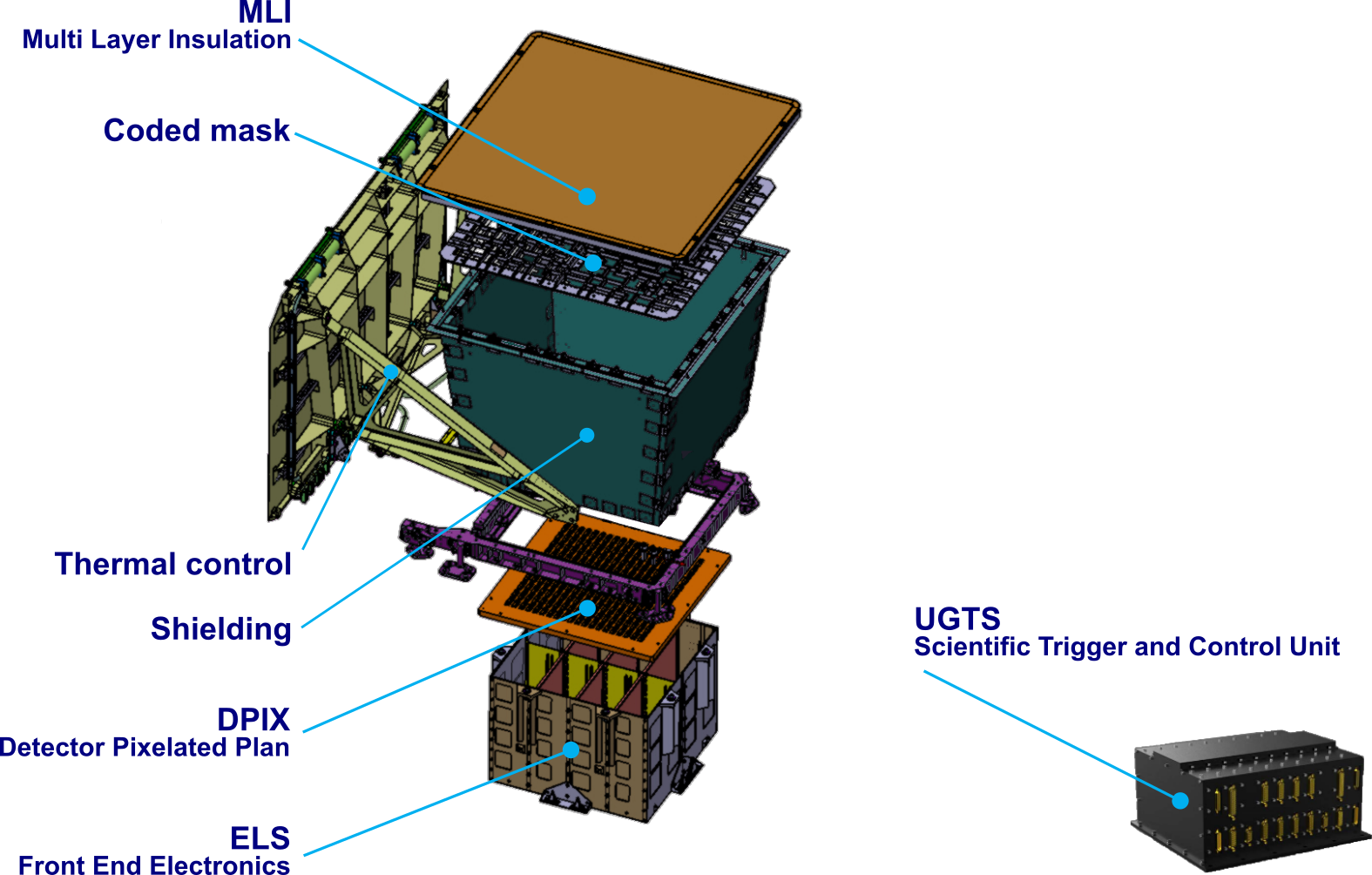}}
\caption{Schematic representation of the ECLAIRs telescope onboard \textit{SVOM}.}
\label{fig:eclairs}
\end{figure}

The ECLAIRs space telescope, mainly dedicated to GRB detection and localisation, is a coded-mask aperture telescope operating in the energy range from $4$ to $150$ keV \citep{takahashi_x-gamma-ray_2014}. With such a low energy threshold, it is particularly well suited for the detection of X-ray rich GRBs and highly redshifted GRBs. Figure \ref{fig:eclairs} shows a schematic representation of ECLAIRs. Its detection plane DPIX is composed of 80$\times$80 CdTe pixels of thickness 1 mm and 4$\times$4 mm$^2$ active surface each, arranged on a 4.5 mm-spaced grid. Its self-supporting tantalum mask has dimensions of 54$\times$54 cm$^2$ and 0.6 mm thickness, providing imaging capabilities up to 120 keV. With its mask-detector distance of 46 cm, the total field of view is 2 sr and the localisation accuracy of sources at detection limit amounts to about 12 arcmin. The onboard data processing, including the trigger algorithms for GRB detection and localisation, is carried out by the Scientific Trigger and Control Unit, called UGTS in French for \textit{Unité de Gestion et de Traitement Scientifique} \citep{schanne_scientific_2013,le_provost_scientific_2013}, based on an FPGA for data preprocessing for the trigger and a Leon3 dual-core CPU running the complete ECLAIRs flight software including the trigger.

As ECLAIRs' tantalum mask is opaque to soft gamma-ray photons, a source emitting such photons will project the shadow of the mask pattern onto the detector, and the image recorded by the detector is called a shadowgram. The sky image is reconstructed from the shadowgram using the mask deconvolution method (as it is, for example, currently used by the IBIS telescope onboard the ESA’s INTEGRAL observatory; see e.g. \cite{goldwurm_integral/ibis_2003} for a description of the method). The deconvolution uses the detected number of counts per pixel $D_{\mathrm{cnt}}$ and, assuming a Poissonian distribution per detector pixel ($D_{\mathrm{var}} = D_{\mathrm{cnt}}$), it produces reconstructed sky images in number of counts ($S_{\mathrm{cnt}}$) and variance ($S_{\mathrm{var}}$) from which an SNR sky image is computed: $S_{\mathrm{SNR}} =  S_{\mathrm{cnt}} / \sqrt{S_{\mathrm{var}}}$. 

To detect GRBs with the ECLAIRs telescope, the UGTS uses two different simultaneously-running trigger algorithms. A count-rate trigger monitors significant rate increases on timescales from $10$ ms up to $20.48$ s followed by the imaging of the sky in the time window of the excess. It is designed to detect short GRBs or long GRBs with short spikes. An image trigger builds $20.48$ seconds-long sky images, which are stacked to systematically form sky images on timescales from $20.48$ s up to $\sim 20$ min, suited for the detection of long and ultra-long GRBs \citep{dagoneau_ultra-long_2020}. These algorithms are implemented in the UGTS (see Fig.~\ref{fig:eclairs}, \citealt{schanne_eclairs_2015, schanne_svom_2019}). The trigger algorithms are foreseen to run in four adjustable energy bands, currently set to: 4--20, 4--50, 4--120, 20--120 keV.

The ECLAIRs telescope is affected by different background components \citep{zhao_influence_2012,mate_simulations_2019}: the Cosmic X-ray Background (CXB) composed of photons from unresolved extra-galactic X-ray sources plus the reflection of the CXB on the Earth's atmosphere and the intrinsic emission of the atmosphere (albedo) caused by the Earth passages in front of the field of view. We will only take into account the CXB while ignoring its reflection and the Earth albedo, since in this study we do not consider the possible presence of the Earth in the field of view of ECLAIRs (and since the CXB dominates the background even in cases with Earth presence). In ECLAIRs, due to the geometry and the large field of view of the instrument, the CXB produces a non-flat close-to-quadratic shape on the detection plane that needs to be removed prior to the deconvolution (see Sec.~\ref{sec:management_cleaning}) in order to ensure the detection of GRBs with a low false-alarm rate and a good sensitivity to faint events. In addition, some known X-ray sources will affect the onboard trigger performances and can be considered as background components. 

\subsubsection{Known X-ray sources in the field of view of ECLAIRs}

Because of the large field of view of ECLAIRs ($\sim$ 2 sr), even if the spacecraft's pointing direction respects the B1 law (see Sec.~\ref{sec:orbit_pointing}) most of the time, this can not be guaranteed during the GRB follow-up after a slew or during a part of the \textit{general program} or target of opportunity observations. As an example, if ECLAIRs points towards a position respecting the B1 law (a Galactic pole for instance), and then a GRB is detected: a slew is performed by the spacecraft and it is very likely that parts of the Galactic plane or bulge, and therefore bright known sources, will enter the field of view. During the long GRB follow-up observation (up to 14 orbits) the bright sources will disturb the trigger algorithms in their search for an other GRB or a source outbreak. Also, during \textit{general program} observations that may occur outside the B1 law, the GRB triggering system is of course enabled. Therefore, some known X-ray sources can be present in the field of view, in particular bright X-ray binaries (such as Sco X-1, Cygnus X-1) or other bright X-ray sources (pulsar wind nebula and/or pulsars, such as the Crab) that are mostly located in the Galactic plane and bulge.  

\subsubsection{Influence on the triggering system}

In the previously described situation, which will represent a large fraction of the overall observing time of ECLAIRs, the onboard trigger algorithms have to deal with the presence of known X-ray sources in the field of view of ECLAIRs that reduce the GRB detection efficiency. Indeed, the known X-ray sources are targets of interest for the \textit{SVOM} \textit{general program} but may be considered as sources of noise for the \textit{core program} and the detection of GRBs by ECLAIRs' triggering system. The known sources in the field of view of ECLAIRs lead to the superposition of different shadows in its detector-shadowgram and to coding noise after deconvolution. In order to enhance the GRB detection capabilities, bright known sources have to be ``cleaned'', i.e. their influence must be reduced in the image reconstruction process. Thus, ECLAIRs' onboard software needs a catalogue of known X-ray sources in order to automatically define the source cleaning strategy and source avoidance during the GRB searches. The detailed cleaning methods used and their performances will be presented in a future paper. 
In the next section, we present the way the catalogue is built including how we generate spectra in order to be able to project the source through a simulator of the ECLAIRs telescope. In Sec.~\ref{sec:simulation} we show the influence of the known X-ray sources on ECLAIRs' noise level and we introduce the way sources will be processed by the onboard computer, and in Sec.~\ref{sec:management_cleaning} the different methods that can be used to clean their contributions. The structure of the onboard catalogue is described in Sec.~\ref{sec:structure}.

\section{Catalogue of sources}

\subsection{Inputs}

We need to build a list of known X-ray sources that may be seen by ECLAIRs in an observation of $\sim$ 20 min duration, which corresponds to the longest exposure time currently foreseen onboard (stacking of 64 sky images with exposures of 20.48 s). Longer exposure mosaics are expected to be constructed on ground by an off-line scientific analysis software. The catalogues of sources needed as input to this off-line software are developed by the team in charge of the ground pipelines and are not discussed in this article. To build the list of sources for the onboard catalogue, we selected all high-energy sources detected by active wide field/all sky monitors. Table \ref{table:missions} gives the list of former and active X-ray wide field instruments.

\begin{table*}
\caption{X-ray wide field instruments (table built with inputs from \citealt{krimm_swift/bat_2013}).}
\label{table:missions}
\centering
\begin{tabular}{|c|c|c|c|c|}
\hline
Mission  & Period of & Energy Range & Source localisation & 20 min sensitivity \\
Instrument & operation & (keV) & accuracy & (3$\sigma$; mCrab)\textsuperscript{a} \\ \hline\hline
\textit{CGRO}/BATSE & 1991--2000 & 20--1800 & > 0.2$\degree$ & 637    \\
\textit{RXTE}/ASM & 1995--2012 & 2--12   & 5 arcmin & 127    \\
\textit{Swift}/BAT & 2004--... & 14--195 & 2.5 arcmin (1$\sigma$) & 136  \\
\textit{Fermi}/GBM & 2008--... & 8--500  & $\approx$ 0.5$\degree$\textsuperscript{b}  & 1274    \\
\textit{MAXI}/GSC & 2009--...  & 2--20   & 1.5$\degree$ & 76    \\
\hline
\end{tabular}
\tablefoot{
\tablefoottext{a}{Assuming the sensitivity is proportional to $\sqrt{t}$, where $t$ is the observing time.}
\tablefoottext{b}{The \textit{Fermi}/GBM localisation accuracy of $\approx$ 0.5$\degree$ corresponds to the one achieved for the monitoring of a catalogue of bright sources with the Earth occultation technique \citep{wilson-hodge_three_2012}. The localisation accuracy is larger for the GRB detection: 3.7 $\degree$ for 90$\%$ of the bursts extending up to $\approx$ 14$\degree$ \citep{connaughton_localization_2015}.}
}
\end{table*}

We need to build a catalogue of sources which permits to estimate the count rate of each source in the various energy bands used by ECLAIRs' onboard software. As a full spectral analysis is beyond the scope of our needs, we use as inputs the data from both \textit{MAXI}/GSC (2--20 keV) and \textit{Swift}/BAT (15--150 keV), which best cover the energy range of ECLAIRs.

\subsubsection{Data for \textit{MAXI}/GSC sources}

\begin{figure}
\resizebox{\hsize}{!}{\includegraphics{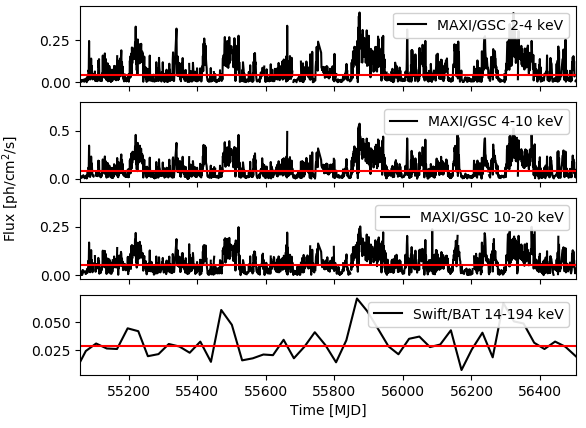}}
\caption{Examples of lightcurves for the source Cen X-3 in the three \textit{MAXI}/GSC bands and in the total \textit{Swift}/BAT band. The red lines indicate the median flux over the lightcurve in each energy band for \textit{MAXI}, and the median of the sum of the fluxes in eight energy bands for \textit{Swift}/BAT.}
\label{fig:lc_src}
\end{figure}

For \textit{MAXI}/GSC \citep{matsuoka_maxi_2009} we downloaded the standard data products\footnote{\url{http://maxi.riken.jp/top/slist.html}} for all of the 425 sources in the \textit{MAXI} list (as of 2019--11--06), including daily binned lightcurves in photons/s/cm$^2$ in three energy bands (2--4, 4--10, 10--20 keV) with associated statistical errors. As an example, we show in Fig.~\ref{fig:lc_src} the lightcurves of the X-ray binary Centaurus X-3 (Cen X-3) obtained in the three \textit{MAXI}/GSC bands.

\subsubsection{Data for \textit{Swift}/BAT sources}

For \textit{Swift}/BAT, we use the BAT 105 months catalogue \citep{oh_105-month_2018} which contains 1632 sources, with source fluxes in units of counts/s/pixel given in eight energy bands covering 14--195 keV. 
The fluxes are extracted from BAT maps built over 105 months and statistical errors are given (hence there is no lightcurve in the eight bands). The Fluxes are converted to photons/s/cm$^2$ using the measured Crab fluxes in the 8 bands given by \cite{baumgartner_70_2013} (Table~2) and assuming a Crab powerlaw spectrum with index -2.15 and normalisation 10.17 ph/s/cm$^2$/keV  \citep{baumgartner_70_2013}. The catalogue also provides a photon powerlaw index over the eight bands. As an example, we show in Fig.~\ref{fig:lc_src} the lightcurve of the X-ray binary Cen X-3 over the complete \textit{Swift}/BAT energy range. The red line gives the median of the sum of the fluxes in the eight bands.

\subsection{Cross-correlation of catalogues and celestial position}

As the \textit{MAXI}/GSC and \textit{Swift}/BAT instruments provide crude positions (in comparison to the ones given by focused instruments such as \textit{Chandra} or \textit{XMM-Newton} in X-rays or even optical positions of their counterparts when they are known), the source positions that will be given in the onboard catalogue are the best positions that are reported in SIMBAD \citep{wenger_simbad_2000}. Most of these come from the Gaia data release \cite{gaia_collaboration_vizier_2018}.

Many sources have various names/identifiers, often associated with different counterparts in various spectral windows. In order to associate a \textit{MAXI} source to a \textit{Swift} one (and vice versa) we use the Sesame name resolver\footnote{http://cds.u-strasbg.fr/cgi-bin/Sesame} for each source to find the main identifier. For instance, Crab has 56 identifiers (M 1, Taurus A, SWIFT J0534.6+2204 ...). Its main identifier is M 1. Few sources are not found by the Sesame name resolver and hence are not associated between \textit{MAXI}/GSC and \textit{Swift}/BAT and in these cases we use the instrumental position. Over the 425 sources from \textit{MAXI}/GSC, 264 have a counterpart in the \textit{Swift}/BAT 105 months catalogue. Sources with no counterpart in one of the two input catalogues are added to the list. This leads to a list of 1793 sources. 

The \textit{Swift}/BAT team, besides providing the 105 months catalogue, also provides a list of sources for the BAT hard X-ray transient monitor \citep{krimm_swift/bat_2013}. This list contains 1026 sources but only 305 are considered as detected by BAT. From these 305 sources, some are already included in the list (from \textit{MAXI} and/or \textit{Swift} side). 44 sources are not present neither in the \textit{MAXI}/GSC list nor in the \textit{Swift}/BAT 105 months catalogue. However, these sources have either a large excess variance ($F_{\mathrm{var}}>3$) or a mean square error larger than the intrinsic sample variance. According to \cite{krimm_swift/bat_2013} they are classified as flaring or outburst sources. That means that they exhibit a high flux over a small time. Such sources should not be considered as disturbing sources for the detection of GRBs and are even sources of interest if detected. This is why we do not project them through ECLAIRs' simulator, as they are not really considered as background sources. If an outburst of such a source is detected onboard, the source will be identified on ground with a search in various catalogues (see Sec.~\ref{sec:src_detection}).

Finally, our list contains 1793 sources (i.e. we found 264 sources present in both  \textit{MAXI}/GSC and \textit{Swift}/BAT catalogues, which are considered only once in our the list of sources). However, it is very likely that ECLAIRs will not detect all of theses sources, even on its longest time scale of 20 min. Based on ECLAIRs' sensitivity that will be determined in Sec.~\ref{sec:sensitivity} and on the source spectra (see Sec.~\ref{sec:spectra}), 53 sources have a flux above the sensitivity limit in the fully coded field of view in 20~min and in at least one of ECLAIRs' four energy bands. In 20~s the number drops to 12 sources. Table \ref{tab:srclist} lists the sources we determined to be bright enough to be detectable by ECLAIRs in 20 min (based on their ``typical SNR'', see Sec.~\ref{sec:sensitivity} and \ref{sec:typical_snr} for the presentation of the selection criteria for this catalogue).

\subsection{Other X-ray sources}
Besides the sources included from the \textit{MAXI}/GSC source list or from the \textit{Swift}/BAT 105 months catalogue, we also investigate if our list of source contains other X-ray sources such as magnetars or type I X-ray bursters. For the magnetars, we cross-match our list of 1793 sources with the list of 30 known magnetars from \cite{olausen_mcgill_2014}. It results that 12 magnetars out of the 30 are included in our list (with XTE J1810-197 among the brightest sources). Concerning the type I X-ray bursters, we cross-match our list with the Multi-INstrument Burst ARchive \citep{galloway_multi-instrument_2020} containing 115 bursts from 85 sources. From this catalogue, 32 sources are missing in our list. Considering the missing magnetars and type I X-ray bursters (50 sources), we do not add them to our list, since they are faint sources (as they are not reported in the \textit{MAXI}/GSC or \textit{Swift}/BAT lists). In the case where the ECLAIRs triggering system detects one of these sources, it will be considered as a GRB candidate and the true identification will be found on ground after a search in catalogues around that position. Newly detected sources (GRB candidates) are temporally included in the onboard catalogue, in order to prevent re-tirggering on them after a first trigger. If desired, typically every week, sources in the onboard catalogue can be removed (or added) by telecommand.

\longtab{
\begin{landscape}
\scriptsize

\begin{longtable}{|c|cccccccc|ccc|ccc|ccc|ccc|}
\caption{\label{tab:srclist}List of the 89 sources in the catalogue, which are the brightest according to their ``typical SNR'' detectable by ECLAIRs in 20 min (ordered according to their median flux in 4--120 keV).}\\
\hline

\multirowcell{2}{Name} & \multirowcell{2}{Ra\\deg} & \multirowcell{2}{Dec\\deg} & \multirowcell{2}{$F_{4-120}$\\ph/s/cm$^2$} & \multirowcell{2}{$\alpha$} & \multirowcell{2}{$\beta$\tablefootmark{a}} & \multirowcell{2}{$E_{\mbox{break}}$\tablefootmark{a}\\keV} & \multirowcell{2}{Type\tablefootmark{b}} & \multirowcell{2}{Class\tablefootmark{c}} & \multicolumn{3}{c|}{4--20 keV} & \multicolumn{3}{c|}{4--50 keV} & \multicolumn{3}{c|}{4--120 keV} & \multicolumn{3}{c|}{20--120 keV} \\ \cline{10-21}
 & & & & & & & & & 20 s\tablefootmark{d} & 20 min\tablefootmark{d} & flag\tablefootmark{e}  & 20 s\tablefootmark{d} & 20 min\tablefootmark{d} & flag\tablefootmark{e}  & 20 s\tablefootmark{d} & 20 min\tablefootmark{d} & flag\tablefootmark{e}  & 20 s\tablefootmark{d} & 20 min\tablefootmark{d} & flag\tablefootmark{e} \\ 
\hline\hline
\endfirsthead

\caption{continued.}\\
\hline

\multirowcell{2}{Name} & \multirowcell{2}{Ra\\deg} & \multirowcell{2}{Dec\\deg} & \multirowcell{2}{$F_{4-120}$\\ph/s/cm$^2$} & \multirowcell{2}{$\alpha$} & \multirowcell{2}{$\beta$\tablefootmark{a}} & \multirowcell{2}{$E_{\mbox{break}}$\tablefootmark{a}\\keV} & \multirowcell{2}{Type\tablefootmark{b}} & \multirowcell{2}{Class\tablefootmark{c}} & \multicolumn{3}{c|}{4--20 keV} & \multicolumn{3}{c|}{4--50 keV} & \multicolumn{3}{c|}{4--120 keV} & \multicolumn{3}{c|}{20--120 keV} \\ \cline{10-21}
 & & & & & & & & & 20 s\tablefootmark{d} & 20 min\tablefootmark{d} & flag\tablefootmark{e} & 20 s\tablefootmark{d} & 20 min\tablefootmark{d} & flag\tablefootmark{e}  & 20 s\tablefootmark{d} & 20 min\tablefootmark{d} & flag\tablefootmark{e}  & 20 s\tablefootmark{d} & 20 min\tablefootmark{d} & flag\tablefootmark{e}  \\ 
\hline\hline
\endhead

\hline
\endfoot

Sco X-1 & 244.98 & -15.64 & 21.157 & -1.74 & -5.89 & 10.59 & LMXB &V& 243.69 & 1949.53 & det & 237.73 & 1901.84 & det & 234.06 & 1872.49 & det & 6.17 & 49.37 & sky \\
GX 5-1 & 270.29 & -25.08 & 1.853 & -1.49 & -6.88 & 11.11 & LMXB &V& 40.96 & 327.65 & det & 37.4 & 299.2 & det & 35.74 & 285.92 & det &  &  &  \\
Crab & 83.63 & 22.01 & 1.793 & -2.19 &  &  & SNR &S& 34.7 & 277.57 & det & 35.8 & 286.42 & det & 35.34 & 282.77 & det & 10.86 & 86.92 & det \\
GX 349+2 & 256.44 & -36.42 & 1.304 & -1.48 & -6.56 & 11.33 & LMXB &V& 30.06 & 240.46 & det & 27.39 & 219.09 & det & 26.13 & 209.05 & det &  &  &  \\
GRS 1915+105 & 288.8 & 10.95 & 1.287 & -2.61 &  &  & LMXB &V& 24.85 & 198.79 & det & 24.22 & 193.79 & det & 23.41 & 187.29 & det & 4.15 & 33.22 & sky \\
GX 17+2 & 274.01 & -14.04 & 1.167 & -1.44 & -5.78 & 10.95 & LMXB &V& 27.29 & 218.31 & det & 24.87 & 198.99 & det & 23.71 & 189.72 & det &  &  &  \\
GX 9+1 & 270.38 & -20.53 & 1.022 & -1.43 & -7.67 & 11.66 & LMXB &S& 24.49 & 195.95 & det & 22.14 & 177.12 & det & 21.06 & 168.5 & det &  &  &  \\
Cyg X-1 & 299.59 & 35.2 & 0.832 & -1.97 &  &  & HMXB &V& 16.77 & 134.18 & det & 18.06 & 144.5 & det & 18.05 & 144.43 & det & 7.16 & 57.26 & det \\
GX 340+0 & 251.45 & -45.61 & 0.822 & -1.08 & -6.11 & 10.43 & LMXB &V& 20.11 & 160.86 & det & 18.23 & 145.81 & det & 17.41 & 139.28 & det &  &  &  \\
Cyg X-2 & 326.17 & 38.32 & 0.724 & -1.87 & -5.84 & 10.29 & LMXB &V& 16.14 & 129.16 & det & 14.62 & 116.96 & det & 13.85 & 110.85 & det &  &  &  \\
GX 13+1 & 273.63 & -17.16 & 0.574 & -1.52 & -6.71 & 10.63 & LMXB &V& 13.68 & 109.48 & det & 12.27 & 98.15 & det & 11.65 & 93.22 & det &  &  &  \\
Sgr X-4& 275.92 & -30.36 & 0.51 & -1.62 & -5.45 & 11.74 & LMXB &V& 12.36 & 98.86 & det & 11.19 & 89.56 & det & 10.65 & 85.23 & det &  &  &  \\
GX 3+1 & 266.98 & -26.56 & 0.415 & -1.51 & -6.81 & 11.82 & LMXB &V& 10.18 & 81.41 & det & 9.29 & 74.36 & det & 8.78 & 70.24 & det &  &  &  \\
4U 1705-440 & 257.23 & -44.1 & 0.405 & -1.42 & -5.99 & 10.45 & LMXB &V& 9.81 & 78.46 & det & 8.77 & 70.13 & det & 8.33 & 66.63 & det &  &  &  \\
Ser X-1 & 279.99 & 5.04 & 0.362 & -3.03 & -6.44 & 17.47 & LMXB &S& 7.3 & 58.42 & det & 6.58 & 52.62 & det & 6.27 & 50.16 & sky &  &  &  \\
GX 9+9 & 262.93 & -16.96 & 0.36 & -1.74 & -7.09 & 12.16 & LMXB &S& 8.65 & 69.17 & det & 7.68 & 61.46 & det & 7.27 & 58.21 & det &  &  &  \\
Cyg X-3 & 308.11 & 40.96 & 0.346 & -1.68 & -3.22 & 18.7 & HMXB &V& 8.05 & 64.38 & det & 8.38 & 67.05 & det & 8.04 & 64.31 & det & 2.29 & 18.34 & sky \\
MAXI J1820+070 & 275.09 & 7.19 & 0.274 & -1.46 &  &  & LMXB && 7.42 & 59.33 & det & 6.62 & 52.99 & det & 6.3 & 50.43 & sky &  &  &  \\
H 1735-444 & 264.74 & -44.45 & 0.272 & -2.34 & -5.97 & 17.27 & LMXB &V& 6.05 & 48.44 & sky & 5.55 & 44.37 & sky & 5.23 & 41.81 & sky &  &  &  \\
Vela X-1 & 135.53 & -40.55 & 0.269 & -0.84 & -3.34 & 20.09 & HMXB &V& 6.01 & 48.1 & sky & 7.06 & 56.46 & det & 6.83 & 54.63 & det & 3.21 & 25.72 & sky \\
SAX J1747.0-2853 & 266.76 & -28.88 & 0.269 & -2.74 &  &  & LMXB &O& 5.57 & 44.56 & sky & 5.33 & 42.68 & sky & 5.05 & 40.42 & sky &  &  &  \\
Sgr A*& 266.42 & -29.01 & 0.265 & -2.51 &  &  & X &S& 5.42 & 43.34 & sky & 5.29 & 42.32 & sky & 5.04 & 40.34 & sky & 0.87 & 6.99 & sky \\
1A 1742-294 & 266.52 & -29.51 & 0.253 & -0.92 & -5.59 & 9.37 & LMXB &V& 6.71 & 53.72 & det & 6.06 & 48.47 & sky & 5.74 & 45.96 & sky &  &  &  \\
H 1730-333& 263.35 & -33.39 & 0.202 & -1.27 & -6.87 & 10.04 & LMXB &O& 5.04 & 40.36 & sky & 4.44 & 35.55 & sky & 4.19 & 33.53 & sky &  &  &  \\
GX 354-0& 262.99 & -33.83 & 0.199 & -1.3 & -3.09 & 8.1 & LMXB &V& 4.83 & 38.64 & sky & 4.48 & 35.86 & sky & 4.32 & 34.6 & sky &  &  &  \\
Cen X-3 & 170.31 & -60.62 & 0.147 & -1.23 & -4.8 & 18.33 & HMXB &V& 3.94 & 31.5 & sky & 3.84 & 30.77 & sky & 3.61 & 28.86 & sky &  &  &  \\
GX 301-2 & 186.66 & -62.77 & 0.138 & -0.33 & -3.27 & 18.46 & HMXB &P& 3.49 & 27.96 & sky & 4.18 & 33.48 & sky & 4.04 & 32.35 & sky & 2.0 & 16.03 & sky \\
SWIFT J0243.6+6124 & 40.92 & 61.43 & 0.11 & -1.06 &  &  & Pulsar && 3.16 & 25.33 & sky & 2.77 & 22.17 & sky & 2.59 & 20.72 & sky &  &  &  \\
4U 1624-490 & 247.02 & -49.21 & 0.107 & -0.86 & -7.74 & 11.51 & LMXB &S& 3.09 & 24.75 & sky & 2.78 & 22.25 & sky & 2.69 & 21.53 & sky &  &  &  \\
GS 1826-238 & 277.37 & -23.8 & 0.104 & -1.99 &  &  & LMXB &V& 2.12 & 16.94 & sky & 2.3 & 18.37 & sky & 2.29 & 18.33 & sky & 0.92 & 7.39 & sky \\
H 1636-536 & 250.23 & -53.75 & 0.093 & -1.5 & -3.03 & 8.6 & LMXB &V& 2.3 & 18.44 & sky & 2.3 & 18.42 & sky & 2.2 & 17.63 & sky &  &  &  \\
4U 1210-64 & 183.27 & -64.92 & 0.092 & -0.69 & -5.57 & 12.32 & HMXB && 2.44 & 19.53 & sky & 2.23 & 17.82 & sky & 2.15 & 17.19 & sky &  &  &  \\
4U 1708-40 & 258.1 & -40.84 & 0.08 & -1.56 & -7.71 & 11.28 & LMXB && 2.09 & 16.69 & sky & 1.83 & 14.62 & sky & 1.77 & 14.13 & sky &  &  &  \\
Perseus Cluster & 49.95 & 41.51 & 0.079 & -3.0 &  &  & Seyfert 2 &S& 1.64 & 13.17 & sky & 1.51 & 12.05 & sky & 1.4 & 11.18 & sky &  &  &  \\
H 0614+091 & 94.28 & 9.14 & 0.078 & -2.54 &  &  & LMXB &V& 1.62 & 12.94 & sky & 1.54 & 12.37 & sky & 1.47 & 11.73 & sky &  &  &  \\
Terzan 2 & 261.89 & -30.8 & 0.077 & -2.31 &  &  & LMXB &V& 1.66 & 13.33 & sky & 1.66 & 13.31 & sky & 1.66 & 13.3 & sky &  &  &  \\
X Per & 58.85 & 31.05 & 0.077 & -2.18 &  &  & HMXB &V& 1.85 & 14.83 & sky & 1.9 & 15.25 & sky & 1.82 & 14.58 & sky &  &  &  \\
SWIFT J1703.9-3753 & 255.99 & -37.84 & 0.075 & -2.62 &  &  & HMXB &V& 1.3 & 10.45 & sky & 2.26 & 18.13 & sky & 2.36 & 18.88 & sky & 2.27 & 18.17 & sky \\
SMC X-1 & 19.27 & -73.44 & 0.073 & -1.66 & -3.35 & 17.75 & HMXB &P& 1.69 & 13.51 & sky & 1.7 & 13.57 & sky & 1.62 & 12.94 & sky &  &  &  \\
4U 1630-472 & 248.51 & -47.39 & 0.071 & -2.47 &  &  & LMXB &O& 1.48 & 11.88 & sky & 1.46 & 11.67 & sky & 1.36 & 10.88 & sky &  &  &  \\
4U 1822-371 & 276.45 & -37.11 & 0.07 & -1.39 & -4.23 & 19.05 & LMXB &S& 2.05 & 16.42 & sky & 2.05 & 16.42 & sky & 1.95 & 15.58 & sky &  &  &  \\
Her X-1 & 254.46 & 35.34 & 0.065 & -0.72 & -4.09 & 20.04 & LMXB &P& 1.66 & 13.31 & sky & 1.77 & 14.13 & sky & 1.71 & 13.68 & sky &  &  &  \\
SWIFT J1753.5-0127 & 268.37 & -1.45 & 0.064 & -1.7 &  &  & LMXB &V& 1.37 & 10.95 & sky & 1.56 & 12.49 & sky & 1.58 & 12.66 & sky &  &  &  \\
OAO 1657-41 & 255.2 & -41.66 & 0.059 & -1.21 & -2.81 & 21.47 & HMXB &V& 1.47 & 11.78 & sky & 1.75 & 13.98 & sky & 1.67 & 13.35 & sky &  &  &  \\
GX 1+4 & 263.01 & -24.75 & 0.058 & -1.06 & -2.36 & 19.94 & Symbiotic* &V& 1.22 & 9.79 & sky & 1.55 & 12.37 & sky & 1.45 & 11.61 & sky &  &  &  \\
4U 1746-37 & 267.55 & -37.05 & 0.058 & -1.58 & -6.01 & 12.25 & LMXB &S& 1.44 & 11.49 & sky & 1.3 & 10.42 & sky & 1.24 & 9.96 & sky &  &  &  \\
4U 1626-67 & 248.07 & -67.46 & 0.057 & -1.39 & -4.21 & 21.05 & LMXB &V& 1.31 & 10.46 & sky & 1.44 & 11.52 & sky & 1.43 & 11.41 & sky &  &  &  \\
4U 1957+115 & 299.85 & 11.71 & 0.056 & -3.17 &  &  & LMXB && 1.1 & 8.84 & sky & 1.06 & 8.51 & sky & 1.01 & 8.07 & sky &  &  &  \\
Cen A & 201.37 & -43.02 & 0.056 & -0.58 & -1.87 & 8.84 & Seyfert 2 &V& 1.34 & 10.7 & sky & 1.51 & 12.1 & sky & 1.51 & 12.13 & sky &  &  &  \\
GX 339-4 & 255.71 & -48.79 & 0.055 & -1.85 &  &  & HMXB &O& 1.24 & 9.93 & sky & 1.38 & 11.03 & sky & 1.44 & 11.53 & sky &  &  &  \\
4U 1254-690 & 194.4 & -69.29 & 0.054 & -1.56 & -6.21 & 11.43 & LMXB &S& 1.29 & 10.29 & sky & 1.25 & 10.02 & sky & 1.23 & 9.87 & sky &  &  &  \\
Ophiuchus Cluster & 258.1 & -23.35 & 0.053 & -1.57 & -4.74 & 10.94 & ClG &S& 1.38 & 11.04 & sky & 1.31 & 10.5 & sky & 1.24 & 9.95 & sky &  &  &  \\
4U 1538-52 & 235.6 & -52.39 & 0.052 & -0.91 & -3.36 & 11.79 & HMXB &V& 1.52 & 12.2 & sky & 1.38 & 11.05 & sky & 1.3 & 10.43 & sky &  &  &  \\
Vela Pulsar& 128.84 & -45.18 & 0.049 & -2.56 &  &  & Pulsar &S& 1.0 & 8.02 & sky & 1.02 & 8.17 & sky & 0.96 & 7.7 & sky &  &  &  \\
4U 1608-52 & 243.18 & -52.42 & 0.049 & -1.27 & -2.4 & 15.0 & LMXB &O& 1.07 & 8.54 & sky & 1.15 & 9.24 & sky & 1.2 & 9.62 & sky &  &  &  \\
Cas A & 350.85 & 58.82 & 0.049 & -3.3 &  &  & SNR &S& 1.1 & 8.8 & sky & 1.01 & 8.11 & sky & 0.96 & 7.65 & sky &  &  &  \\
HETE J1900.1-2455 & 285.04 & -24.92 & 0.048 & -2.13 &  &  & LMXB &V& 1.22 & 9.76 & sky & 1.21 & 9.72 & sky & 1.14 & 9.14 & sky &  &  &  \\
4U 1543-624 & 236.98 & -62.57 & 0.046 & -1.66 & -5.53 & 11.39 & LMXB &S& 1.11 & 8.89 & sky & 0.99 & 7.95 & sky & 0.96 & 7.72 & sky &  &  &  \\
MAXI J1910-057 & 287.59 & -5.8 & 0.045 & -0.94 &  &  & LMXB &O&  &  &  & 0.99 & 7.94 & sky & 1.24 & 9.93 & sky & 1.21 & 9.66 & sky \\
H 1822-000 & 276.34 & -0.01 & 0.044 & -1.7 & -6.51 & 12.45 & LMXB &S& 0.87 & 6.94 & sky &  &  &  & 0.83 & 6.65 & sky &  &  &  \\
XTE J1701-407 & 255.43 & -40.86 & 0.043 & -2.57 &  &  & LMXB &V& 0.82 & 6.59 & sky & 0.91 & 7.29 & sky & 0.89 & 7.13 & sky &  &  &  \\
MAXI J1631-479 & 247.81 & -47.81 & 0.04 & -1.38 &  &  & X && 1.29 & 10.32 & sky & 1.17 & 9.38 & sky & 1.08 & 8.64 & sky &  &  &  \\
Coma Cluster & 194.95 & 27.98 & 0.039 & -2.87 &  &  & Clg&& 0.84 & 6.72 & sky &  &  &  &  &  &  &  &  &  \\
Aql X-1 & 287.82 & 0.58 & 0.037 & -1.2 & -2.82 & 10.1 & LMXB &O& 0.86 & 6.93 & sky & 0.93 & 7.42 & sky & 0.95 & 7.57 & sky &  &  &  \\
EXO 2030+375 & 308.06 & 37.64 & 0.037 & -0.75 & -2.89 & 18.65 & HMXB &P&  &  &  & 0.86 & 6.87 & sky & 0.82 & 6.56 & sky &  &  &  \\
Terzan 5 & 267.02 & -24.78 & 0.037 & -1.71 & -5.32 & 10.85 & Cl* &O& 0.86 & 6.91 & sky & 0.88 & 7.03 & sky & 0.89 & 7.09 & sky &  &  &  \\
1E1145.1-6141 & 176.87 & -61.95 & 0.035 & -1.46 & -2.85 & 17.34 & HMXB &V& 0.98 & 7.87 & sky & 1.07 & 8.6 & sky & 1.05 & 8.43 & sky &  &  &  \\
4U 1907+09 & 287.41 & 9.83 & 0.034 & -1.77 & -4.04 & 20.35 & HMXB &V& 0.92 & 7.4 & sky & 0.96 & 7.65 & sky & 0.95 & 7.57 & sky &  &  &  \\
4U 1954+319 & 298.93 & 32.1 & 0.033 & -1.85 & -3.04 & 19.62 & HMXB &O&  &  &  & 0.83 & 6.61 & sky &  &  &  &  &  &  \\
MAXI J1813-095 & 273.25 & -9.5 & 0.033 & -0.85 &  &  & X && 0.89 & 7.17 & sky &  &  &  &  &  &  &  &  &  \\
LMC X-2 & 80.12 & -71.96 & 0.033 & -1.72 & -7.27 & 11.73 & LMXB && 0.94 & 7.54 & sky &  &  &  &  &  &  &  &  &  \\
EXO 1722-363 & 261.3 & -36.28 & 0.033 & -2.07 & -3.07 & 22.0 & HMXB &V&  &  &  & 0.81 & 6.51 & sky &  &  &  &  &  &  \\
XTE J1550-564 & 237.74 & -56.48 & 0.032 & -0.84 &  &  & HMXB &&  &  &  & 0.83 & 6.63 & sky & 0.83 & 6.67 & sky &  &  &  \\
A 0535+262 & 84.73 & 26.32 & 0.032 & -0.87 & -3.07 & 22.83 & HMXB &O& 0.84 & 6.74 & sky & 1.02 & 8.14 & sky & 0.97 & 7.74 & sky &  &  &  \\
EXO 1846-031 & 282.32 & -3.06 & 0.032 & -1.25 &  &  & LMXB && 0.86 & 6.88 & sky &  &  &  &  &  &  &  &  &  \\
SWIFT J1658.2-4242 & 254.55 & -42.7 & 0.031 & -1.57 &  &  & X && 0.99 & 7.94 & sky & 0.92 & 7.39 & sky & 0.92 & 7.36 & sky &  &  &  \\
SWIFT J1728.9-3613 & 262.23 & -36.24 & 0.029 & -1.13 &  &  & X && 1.02 & 8.2 & sky & 0.92 & 7.39 & sky & 0.87 & 6.96 & sky &  &  &  \\
4U 2206+543 & 331.98 & 54.52 & 0.029 & -1.09 & -2.58 & 8.11 & HMXB &V& 0.9 & 7.17 & sky & 0.87 & 6.99 & sky & 0.83 & 6.64 & sky &  &  &  \\
SMC X-2 & 13.64 & -73.68 & 0.029 & -1.18 &  &  & HMXB && 1.05 & 8.45 & sky & 1.01 & 8.08 & sky & 0.92 & 7.38 & sky &  &  &  \\
MCG -05-23-016 & 146.92 & -30.95 & 0.028 & -1.99 &  &  & Seyfert 2 &S& 1.11 & 8.89 & sky & 1.14 & 9.15 & sky & 1.11 & 8.86 & sky &  &  &  \\
H 1743-322 & 266.56 & -32.23 & 0.027 & -1.94 &  &  & LMXB &O& 0.82 & 6.55 & sky & 0.84 & 6.74 & sky & 0.87 & 6.98 & sky &  &  &  \\
IGR J17379-3747 & 264.5 & -37.77 & 0.027 & -1.03 &  &  & HMXB&& 0.83 & 6.69 & sky & 0.82 & 6.55 & sky &  &  &  &  &  &  \\
GX 304-1 & 195.32 & -61.6 & 0.026 & -0.77 & -3.2 & 19.48 & HMXB &O& 0.81 & 6.5 & sky & 0.87 & 6.99 & sky &  &  &  &  &  &  \\
XTE J1810-197 & 272.46 & -19.73 & 0.026 & -1.63 &  &  & Pulsar &O& 0.81 & 6.51 & sky &  &  &  &  &  &  &  &  &  \\
MAXI J0511-522 & 77.75 & -52.2 & 0.025 & -0.92 &  &  & X && 0.92 & 7.41 & sky &  &  &  &  &  &  &  &  &  \\
AM Her & 274.06 & 49.87 & 0.025 & -2.49 &  &  & AMHer &S& 1.01 & 8.1 & sky & 0.92 & 7.41 & sky & 0.9 & 7.19 & sky &  &  &  \\
1ES 1011+496 & 153.77 & 49.43 & 0.023 & -0.74 &  &  & BLLac && 0.89 & 7.16 & sky & 0.85 & 6.84 & sky & 0.84 & 6.75 & sky &  &  &  \\
GS 1843-02 & 282.07 & -2.42 & 0.022 & -1.06 &  &  & HMXB &O& 0.85 & 6.79 & sky &  &  &  &  &  &  &  &  &  \\
GRO J1655-40 & 253.5 & -39.85 & 0.016 & -1.33 & -3.26 & 9.25 & HMXB &O& 0.95 & 7.61 & sky & 0.83 & 6.65 & sky &  &  &  &  &  &  \\

\end{longtable}

\tablefoot{
\tablefoottext{a}{$\beta$ and $E_{\mbox{break}}$ are given only for sources whose best spectral model is a broken powerlaw.}
\tablefoottext{b}{Type acronyms: ClG (cluster of galaxies), GlCl (globular cluster), HMXB (high mass X-ray binary), LMXB (low mass X-ray binary), SNR (supernova remnant), XB (other X-ray binary), X (other X-ray source).}
\tablefoottext{c}{Classes are from \cite{krimm_swift/bat_2013}: S (steady), V (variable), P (periodic), O (outburst), sources with no class are not detected by the BAT monitor or detected after 2013.}
\tablefoottext{d}{Typical SNR in 20 s or 20 min sky image (see Sec.~\ref{sec:typical_snr}).}
\tablefoottext{e}{Flag specifying the strategy for the source management in the trigger algorithm : det = source is fit and subtracted in the detector shadowgram prior to sky reconstruction, sky = source position is masked in the reconstructed sky image (see Sec.~\ref{sec:structure}).}
}

\end{landscape}
}

\subsection{Spectra}
\label{sec:spectra}

Now that we have a list of sources we can generate the spectra for all 1793 sources, using the \textit{MAXI}/GSC and/or \textit{Swift}/BAT 105 months data. Spectra are needed to be able to simulate the sources in ECLAIRs' model, and to evaluate their influence on the background level (see Sec.~\ref{sec:simulation}). To build those spectra, we proceed as follows.

\begin{enumerate}
    \item If the source has \textit{MAXI}/GSC and \textit{Swift}/BAT data, a joint broadband fit is performed with a single powerlaw (Eq. \ref{eq:powerlaw}) and a broken powerlaw (Eq.~\ref{eq:brokenpl}):
    
    \begin{equation}
        \label{eq:powerlaw}
        N(E) = K \cdot E^{\alpha}
    \end{equation}
    
    \begin{equation}
        \label{eq:brokenpl}
        N(E) = \left\{
        \begin{array}{ll}
            K \cdot E^{\alpha} & \mbox{if } E \leq E_{\mbox{break}}\\
            K \cdot E^{\beta} \cdot E_{\mbox{break}}^{\alpha-\beta} & \mbox{if } E > E_{\mbox{break}}
        \end{array}
        \right.
    \end{equation}
    
    where $N(E)$ is the photon flux density (i.e. the number of photons per unit energy bandwidth per unit area per second). K stands for the spectrum normalisation (photons/cm$^2$/s/keV at 1 keV), $\alpha$ and $\beta$ are the photon indexes. In the case of a broken powerlaw, $E_{\mbox{break}}$ is the break energy at which the slope changes. The fitted fluxes are the median fluxes (preferred to the mean, in order to get a flux which is less dependent on potential outbreaks of faint sources) of the three \textit{MAXI}/GSC sub-band lightcurves plus the fluxes of the eight \textit{Swift}/BAT sub-bands (given in the source pulse height analyser file). The energy corresponding to each flux point is set to the centre of its sub-band. Out of both performed fits, we choose to keep the powerlaw (instead of the broken powerlaw) if one of the following is true:
    
    \begin{itemize}
        \item the summed squared error of the single powerlaw is smaller than the one of the broken powerlaw;
        \item the break energy is below 8 keV or above 80 keV;
        \item the slope is steeper at low energy than at high energy ($|\alpha| > |\beta|$);
        \item the difference of the two photon indexes of the broken powerlaw is smaller than 1 (it leads to an error on the flux of 10$\%$ or less for the 17 sources affected by this constraint).
    \end{itemize}

    \item If the source has only \textit{MAXI}/GSC data, a fit is performed with a single powerlaw (Eq. \ref{eq:powerlaw}). The fitted fluxes are the median fluxes in the three \textit{MAXI}/GSC sub-band lightcurves.
    
    \item If the source has only \textit{Swift}/BAT data, the spectrum is the single powerlaw (Eq. \ref{eq:powerlaw}) given by the BAT 105 month catalogue \citep{oh_105-month_2018}.
\end{enumerate}

In a last step, if a source is detected by \textit{MAXI}/GSC, its spectrum normalisation is corrected to ensure that the integration of our spectrum reaches the median flux of \textit{MAXI}/GSC in 4--20 keV. It allows to correct for a possible offset between BAT and GSC spectra. This offset may arise after fitting the fluxes computed from different time windows in the two instruments for sources that are known to vary within typical scales of days/months.

The absorption is not taken into account since ECLAIRs will operate above 4 keV and all sources in the catalogue have an HI column density $N_{\mathrm{H}} < 2.27 \cdot 10^{22}$ atoms/cm$^{2}$ with a mean of $ 2.63 \cdot 10^{21}$ atoms/cm$^{2}$ (see Sec.~\ref{sec:population}). Compared to a non-absorbed powerlaw ($\alpha = -2$, $K = 10$), taking into account this mean $N_{\mathrm{H}}$ value reduces the flux by 0.6 $\%$ in 4--20 keV (the influence is even smaller on larger energy bands).

\begin{figure}
\resizebox{\hsize}{!}{\includegraphics{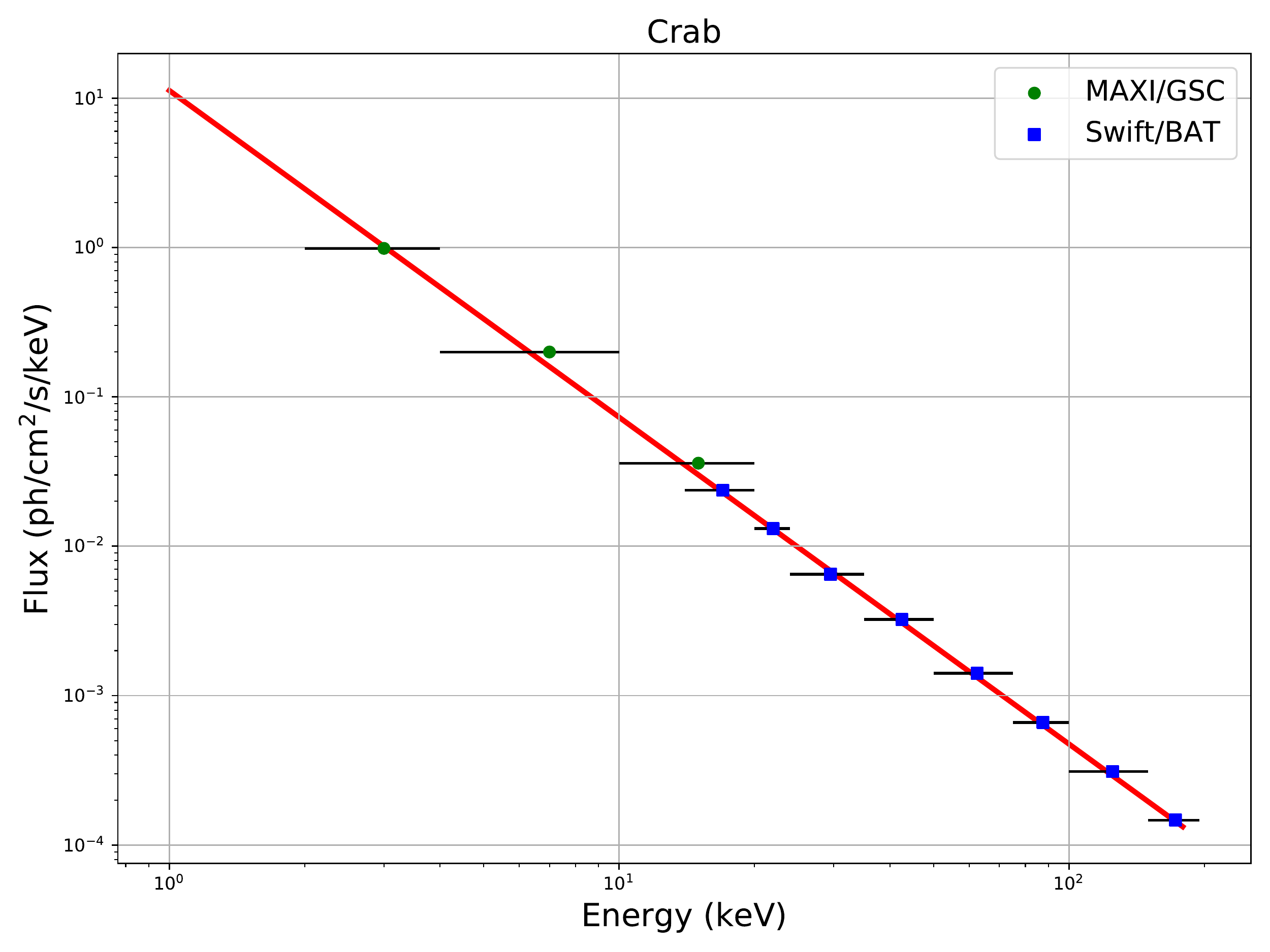}}
\caption{Example of spectrum for Crab obtained by our procedure.}
\label{fig:spCrab}
\end{figure}

In the case of the well-known Crab, this procedure leads to the reconstructed spectrum shown in Fig.~\ref{fig:spCrab}. We find the best model being a single powerlaw with a normalisation $K=11.2\pm1.1$ ph/s/cm2/keV (at 1 keV, 1$\sigma$) and a photon index $\alpha=-2.19\pm0.03$ (1$\sigma$). This result is very close to the one usually reported in the literature \citep{toor_crab_1974,weisskopf_calibrations_2010}.

\begin{figure}
\resizebox{\hsize}{!}{\includegraphics{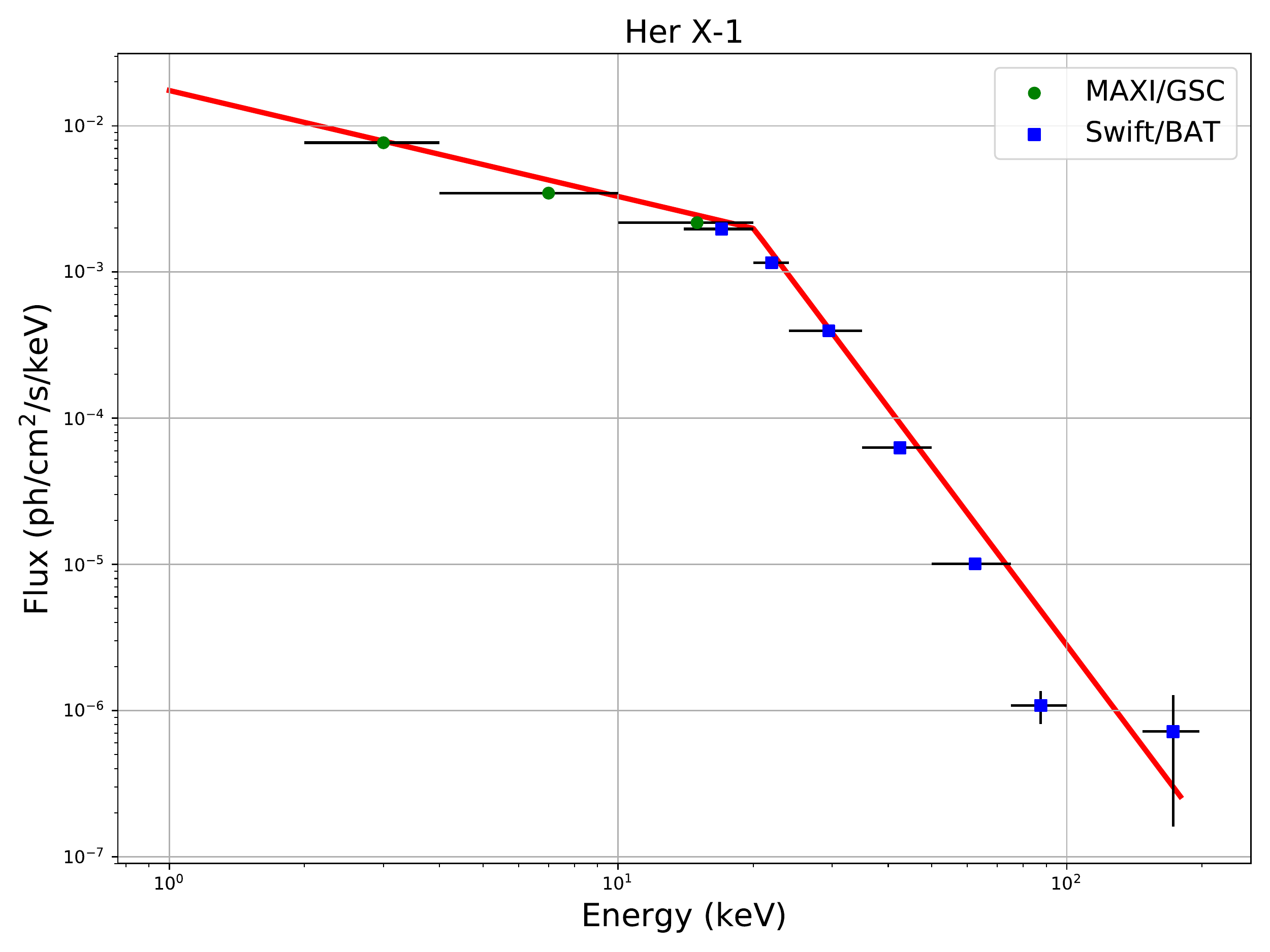}}
\caption{Example of spectrum for Her X-1 obtained by our procedure.}
\label{fig:spHer}
\end{figure}

Figure \ref{fig:spHer} gives an example of a spectrum for the source Hercules X-1 (Her X-1), whose best fit model is a broken powerlaw with $\alpha=-0.72\pm0.21$ (1$\sigma$), $\beta=-4.09\pm0.18$ (1$\sigma$) and $E_{\mbox{break}}=20.04\pm0.46$ keV (1$\sigma$). These values agree with the ones from \cite{dal_fiume_broad-band_1998}. Note, however, that these authors have a best model consisting of an exponentially cut-off powerlaw with $\alpha=-0.884\pm0.003$ (90$\%$ C.L.) and a cut-off energy of $24.2\pm0.2$ keV. We also estimate an approximate slope of -4.2 from the spectra shown in their Fig. 1 (right).

Note that the reduced $\chi^2$ are large ($3.8\cdot 10^4$ for Crab and $6.3\cdot 10^2$ for Her X-1) which may be caused by the fact that the sources are bright and have small statistical errors and also by the coarse energy binning\footnote{The flux in an energy bin is distributed across the entire bin width with a specific law and not just in its centre as we assumed it. Therefore, the error on the flux in the centre of the bin depends on the statistical error and on a systematic error based on the bin width. Taking this error into account should help to reduce the $\chi^2$.}. This effect has already been observed by \cite{baumgartner_70_2013} for the fit of BAT spectra with a simple powerlaw. Also, systematic errors are not taken into account. 

Also, some sources are known to exhibit some spectral changes (e.g. Sco X-1, GRS 1915+105, Cyg X-3, Cyg X-1 ...). These changes are related to different behaviours of the accretion disk around the compact object in X-ray binaries. In our study, as the spectra are built over a long time of observation, the best models we obtain for these sources do not represent one particular state. We remind that we only need spectra to be able to evaluate the influence of the sources on the background level of ECLAIRs and a detailed spectral analysis is beyond the scope of our needs. Moreover the bright sources will be cleaned regardless of their states (see Sec.~\ref{sec:management_cleaning}). 

As a result, 1698 sources can be well modelled with a single powerlaw spectrum and 95 sources are described by a broken powerlaw. Table \ref{tab:srclist} also gives the spectral parameters of the brightest sources. The full table giving the general properties, the data in the spectral bands from \textit{MAXI}/GSC and \textit{Swift}/BAT, and the spectral fit we performed, for all the 1793 sources, is available at the CDS.

In the simulations presented in section \ref{sec:simulation}, the sources will be projected through ECLAIRs' model, using the photon flux obtained by the integration of the spectrum in a maximum energy range of 4 to 150 keV. Photons from sources without \textit{Swift}/BAT counterpart will be rejected above 20 keV, while photons from sources without \textit{MAXI}/GSC counterpart (present only in the BAT 105 month catalogue) will be rejected below 14 keV. The sources only present in the \textit{Swift}/BAT hard X-ray transient monitor list will not be projected.

\subsection{Population statistics}
\label{sec:population}

\begin{figure*}
\centering
\includegraphics[width=17cm]{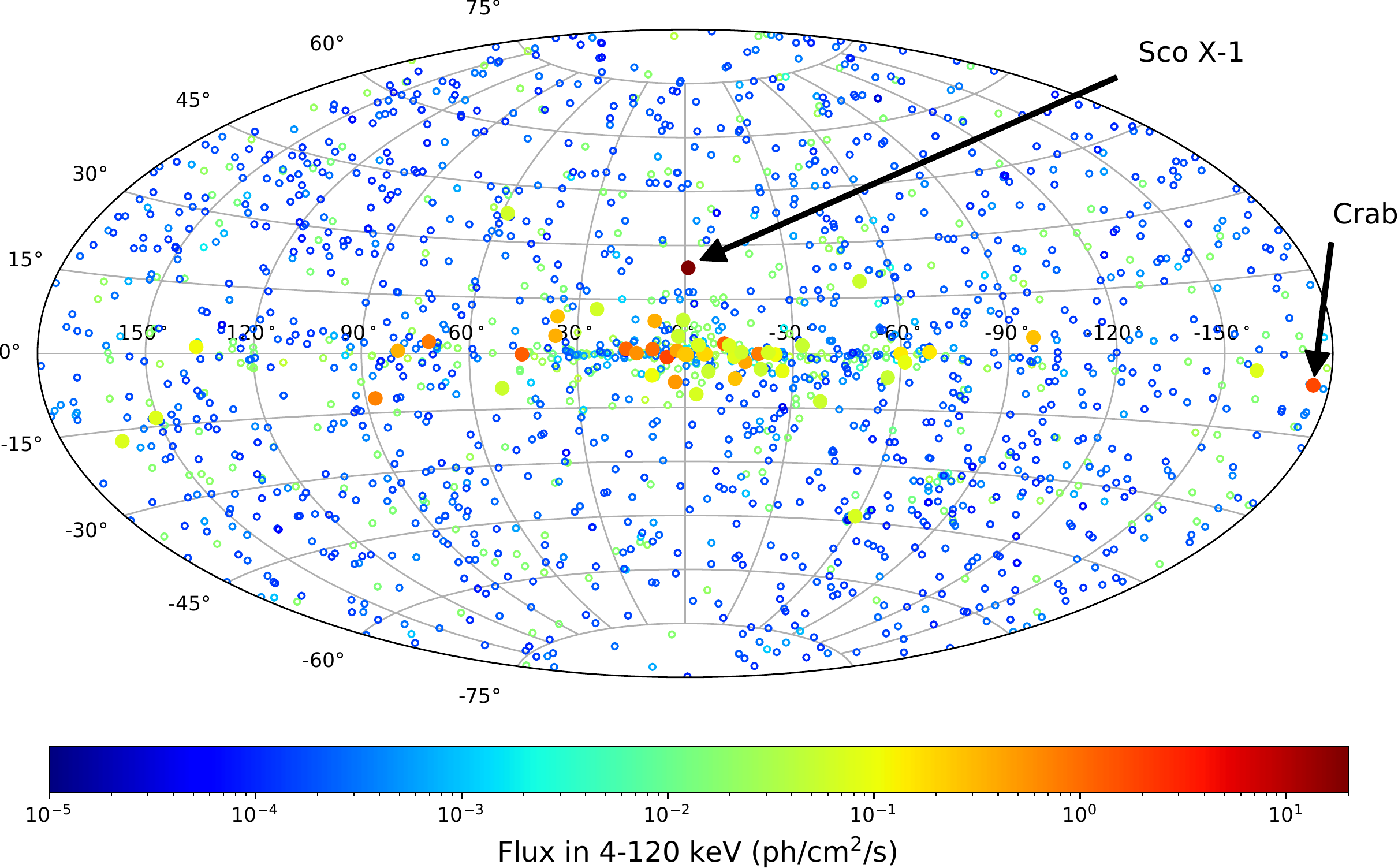}
\caption{Distribution of the sources of our catalogue on the sky in Galactic coordinates. Empty circles correspond to the full catalogue and plain circles correspond to the sources that should be visible in 20 min according to the expected ECLAIRs sensitivity. The colour gives the flux in 4--120 keV in ph/cm$^2$/s.}
\label{fig:map}
\end{figure*}

In this section we provide some statistics on the catalogue we have built. First, the Fig.~\ref{fig:map} shows the distribution of the sources in the sky. The plain circles, which depict the 53 brightest sources of the catalogue that ECLAIRs should be able to detect in 20 min images (according to its sensitivity, see \ref{sec:sensitivity}), are clustered in the Galactic plane. These sources are mostly accreting X-ray binaries with either a neutron star or a black hole as the primary compact object.

\begin{figure}
\resizebox{\hsize}{!}{\includegraphics{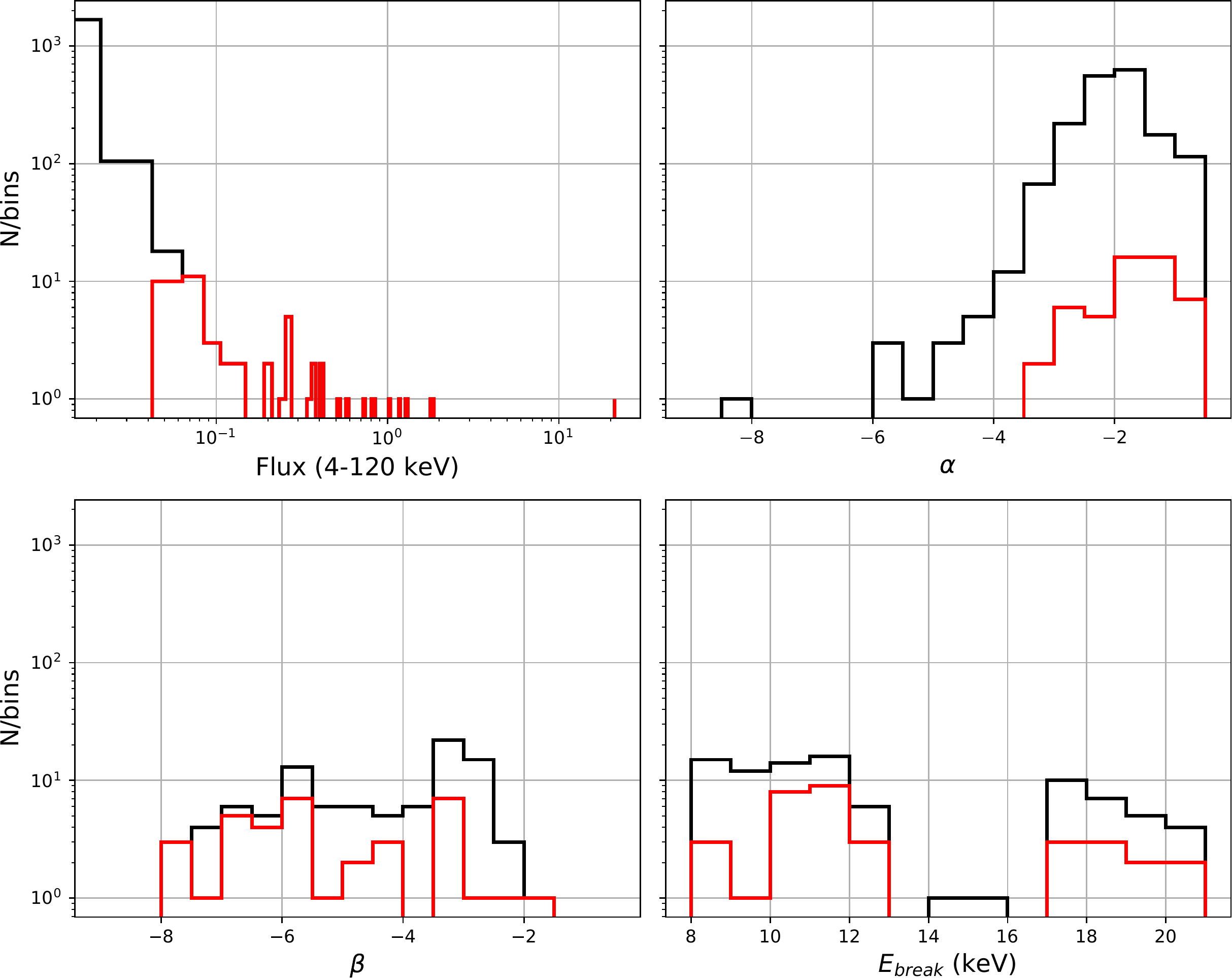}}
\caption{Distribution of the spectral parameters. Top left: 4--120 keV flux (ph/cm$^2$/s). Top right: photon index $\alpha$. Bottom left: photon index $\beta$ (for the sources with a broken powerlaw spectrum). Bottom right: break energy in keV: $E_{\mbox{break}}$ (for the sources with a broken powerlaw spectrum). Black histograms: full catalogue, red histograms: brightest 53 sources.}
\label{fig:stat_spectrum}
\end{figure}

\begin{figure}

\resizebox{\hsize}{!}{\includegraphics{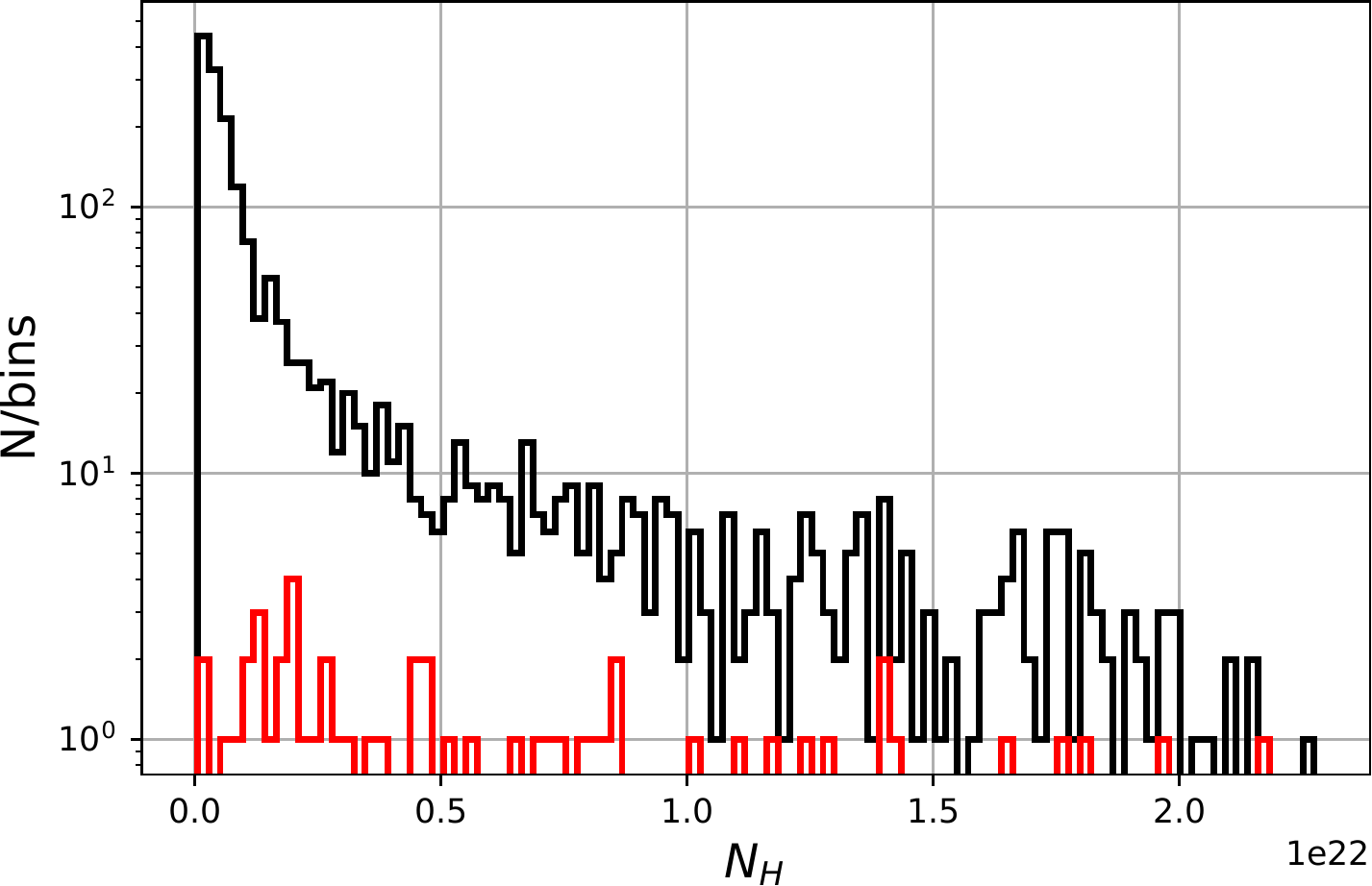}}
\caption{Distribution of the hydrogen column density (in units of 10$^{22}$ atoms/cm$^2$) $N_{\mbox{H}}$ at the source positions retrieved from \cite{bekhti_hi4pi_2016}. Black: full catalogue, red: brightest 53 sources.}
\label{fig:nh}
\end{figure}

Figure \ref{fig:stat_spectrum} gives the distributions of the flux in the 4--120 keV spectral range and the ones of the spectral parameters obtained from the fits to our reconstructed spectra (Sec.~\ref{sec:spectra}) : $\alpha$, $\beta$ and $E_{\mbox{break}}$. The mean value of the powerlaw index is $\alpha=-2.01\pm0.65~(1\sigma)$. In the break energy histogram (Fig.~\ref{fig:stat_spectrum} bottom right), two groups are visible at $E_{\mbox{break}} \sim 10$ keV and $E_{\mbox{break}} \sim 18$ keV. The latter corresponds to breaks between \textit{MAXI}/GSC and \textit{Swift}/BAT energy bins and is consistent with a broken powerlaw (see Fig.~\ref{fig:spHer}). The former group has a break value only within the \textit{MAXI}/GSC spectral range. In this population, the low value of the energy break could, in part, phenomenologically mimic the effect of absorption that we have not taken into account. The absorption indeed has a stronger effect on the low energy photons and therefore tends to bend the low part of the spectra towards smaller values. The resultant spectra are therefore harder: their photon index is smaller (in absolute value) below the energy where the absorption no longer affects the radiation (taken as $E_{\mbox{break}}$). Figure \ref{fig:nh} shows the distribution of the hydrogen column density at the sources' position.

Table \ref{tab:pop_stat} gives a summary of the minimum, maximum, mean and standard deviation of each of the previous values.

\begin{table*}
\caption{Summary of the catalogue statistics.}
\label{tab:pop_stat}
\centering
\begin{tabular}{|c|c|c|c|c|c|c|}
\hline
& & \multirowcell{2}{$F_{4-120}$\\(ph/cm$^2$/s)} &  \multirowcell{2}{$\alpha$} &  \multirowcell{2}{$\beta$} & \multirowcell{2}{$E_{\mbox{break}}$\\(keV)} & \multirowcell{2}{$N_{\mbox{H}}$\\(atoms/cm$^2$)} \\
& & & & & & \\ \hline \hline
\multirowcell{4}{All sources} & min & 0.00 & -8.25 & -7.74 & 8.05 & 6.08e+19 \\
 & max  & 21.16 & -0.02 & -1.87 & 22.00 & 2.27e+22 \\
 & mean & 0.03  & -2.01 & -4.46 & 13.11 & 2.63e+21 \\
 & std  & 0.51  & 0.65  & 1.58  & 4.18  & 4.31e+21 \\
\hline
\multirowcell{4}{Brightest\\53 sources} & min & 0.05 & -3.17  & -7.74 & 8.10 & 1.54e+20 \\
 & max  & 21.16 & -0.33 & -1.87 & 21.47 & 2.17e+22 \\
 & mean & 0.74  & -1.61 & -5.18 & 13.64 & 6.48e+21 \\
 & std  & 2.87  & 0.60  & 1.66  & 4.15  & 5.64e+21 \\
\hline                                                        
\end{tabular}
\end{table*}

Table \ref{tab:srctype} gives the source types obtained from the BAT 105 months catalogue \citep{oh_105-month_2018}, or from the Simbad database \citep{wenger_simbad_2000} if unknown from BAT, for the full catalogue and for the 53 brightest sources (according to the sensitivity, see Sec.~\ref{sec:sensitivity}). Most of the sources in the catalogue are active galaxies or X-ray binaries. The brightest sources are mainly low-mass X-ray binaries.

\begin{table}
\caption{Summary of the source types.}
\label{tab:srctype}
\centering
\begin{tabular}{|c|c|c|}
\hline
\multirowcell{2}{Type} & \multirowcell{2}{All sources} & \multirowcell{2}{Brightest\\53 sources} \\
 & & \\ \hline\hline
AGN & 1127 & 2 \\ 
Galaxy clusters/groups & 52 & 1 \\ 
Others galaxies & 9 & 0 \\ 
 \hline 
HMXB & 135 & 12 \\ 
LMXB & 132 & 34 \\ 
Others XRB & 13 & 0 \\ 
\hline 
Stars & 114 & 1 \\ 
Nov\ae & 29 & 0 \\ 
Pulsars & 28 & 1 \\ 
SNR & 11 & 1 \\  
HII/clouds & 4 & 0 \\ 
Star clusters & 1 & 0 \\ 
\hline
Unknown & 180 & 1 \\ 
\hline
\end{tabular}
\end{table}

\section{Simulations through ECLAIRs}
\label{sec:simulation}

In this section, we present the simulations we have performed in order to estimate the influence of the  catalogue sources on ECLAIRs' background level. First we determine the sensitivity of ECLAIRs with respect to the sources in Sec.~\ref{sec:sensitivity}, then we simulate the sources through ECLAIRs' model in Sec.~\ref{sec:simu_source}. For the different simulations, we use our ray-tracing simulation software that propagates photons one by one through ECLAIRs' coded mask. The ray-tracing takes into account the energy redistribution which includes the transparency of the mask and the efficiency of the detector according to the energy of the incident photons. Additionally, in the following simulations, CXB photons are also propagated through ECLAIRs with a spectrum given by \cite{moretti_new_2009} (we use the CXB spectrum obtained from \textit{Swift}/XRT and BAT data in the 1.5--200 keV band, described by two smoothly-joined power laws with an energy break at 29 keV and two photon indices of 1.40 and 2.88) and a small internal background of 0.003 counts/s/cm$^2$/keV with a flat spectrum is added \citep{sizun_synthesis_2011} as an estimated mean contribution of particles over the orbit. In this study we assume that the field of view is completely free of the Earth.

\subsection{Sensitivity of ECLAIRs}
\label{sec:sensitivity}

\begin{figure*}
\centering
\includegraphics[width=17cm]{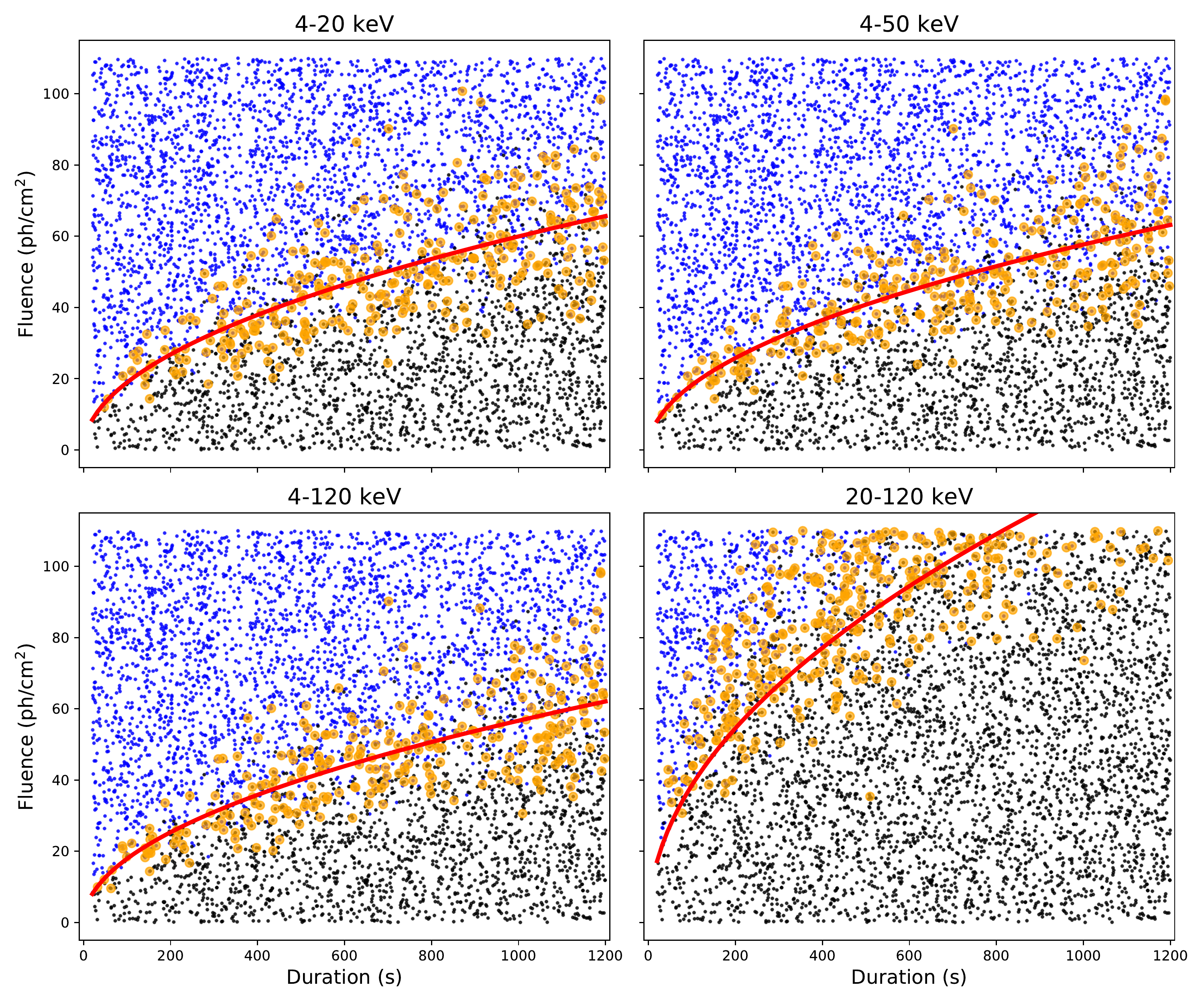}
\caption{ECLAIRs' sensitivity in different energy bands according to the exposure time. Blue points indicate sources with SNR $\geqslant$ 6.5, black points indicate sources with SNR $<$ 6.5. Orange points indicate sources with 6 $\leqslant$ SNR $<$ 7 used to fit the sensitivity law (red curve).}
\label{fig:sensitivity}
\end{figure*}

In our catalogue there are 1793 sources detected by \textit{MAXI}/GSC or \textit{Swift}/BAT (or both). As \textit{MAXI}/GSC and \textit{Swift}/BAT have a better sensitivity than ECLAIRs, many of these sources will be too faint to be seen by ECLAIRs in 20 min. To obtain the sub-catalogue of the bright sources that will be seen by ECLAIRs in 20 min, we first need to estimate ECLAIRs' sensitivity in each of the overlapping trigger energy bands (``energy strips'') foreseen on board (4--20, 4--50, 4--120, 20--120 keV). To do so, we simulated 5000 dummy sources in the fully-coded field of view with uniformly drawn fluence between 0 and 110 ph/cm$^2$ over an exposure time uniformly drawn between 20 s and 20 min. The maximum fluence of 110 ph/cm$^2$ was chosen in order to fill the fluence-duration plane of our simulation with enough sources to be able to derive the detection sensitivity (over background) close to the maximum duration of 20 min. The dummy sources were ray-traced one by one, with the CXB and the internal background added. The source spectrum is assumed to be a simple powerlaw of index $-2$ which corresponds to the mean photon index in our catalogue (see Sec.~\ref{sec:population}). The shadowgrams are cleaned before the deconvolution by fitting and subtracting a quadratic-shape model of the CXB. After the cleaning and the deconvolution, the source SNR is obtained at the source position in the sky image. For each energy band, a square root law $S = S_1\cdot\sqrt{t}$ is fitted ($S$ denotes the fluence over the observing time $t$, and $S_1$ the 1 second fluence at the detection limit) using the sources close to a detection threshold of SNR$=$6.5 \citep{schanne_presentation_2009}. This value corresponds to the threshold in the SNR sky images required to claim the detection of a new source, while ensuring a low false detection rate (less than one per day). The results for the four energy bands are shown in Fig.~\ref{fig:sensitivity} and Table~\ref{tab:sensitivity}.

The sensitivity and associated errors in the three firsts bands (which start at 4 keV) are very close ($\approx 1.8$ ph/cm$^2$/s, see Table~\ref{tab:sensitivity} for precise values). When the high energy bound increases, the ratio between the number of photons of the source and the number of photons of the CXB decreases. In other words, the sources-to-background relative contribution decreases with energy. Thus above 20 keV, more photons are required, and therefore also a higher fluence, to reach the detection threshold. This effect is shown in the bottom right plot of Fig.~\ref{fig:sensitivity}. The sensitivity therefore drops significantly above 20 keV. 

\begin{table}
\caption{ECLAIRs' sensitivity in different energy bands for detection with SNR>6.5. $S_1$ is the 1 second sensitivity and the sensitivity follows $S(t) = S_1\cdot\sqrt{t / 1 \mathrm{s}}$.}
\label{tab:sensitivity}
\centering
\begin{tabular}{|c|c|c|}
\hline
Energy range & $S_1$ (ph/cm$^2$) & err (ph/cm$^2$, $1\sigma$)  \\\hline\hline
4--20 keV & 1.895 & 0.026 \\
4--50 keV & 1.823 & 0.026 \\
4--120 keV & 1.792 & 0.024 \\
20--120 keV & 3.855 & 0.045 \\
\hline
\end{tabular}
\end{table}

We remark here that the sensitivity computed with our methods should not be taken as absolute sensitivity values for ECLAIRs, and should therefore be considered with the following caveats. First, our sensitivity values are calculated only in the fully-coded field of view (0.15 sr) and they will decrease in the total field of view. Then, they are strongly dependent on the assumed background shape and level. This is why the measured sensitivity during flight operations may differ from the one we give here.

With these calculated sensitivities (Table~\ref{tab:sensitivity}) we can extract the list of the known sources which may be detected in one of the four considered energy bands in a maximum accumulation time of 20 min. This list, or sub-catalogue, is composed of 53 sources and was used in the previous sections to provide some statistics on the brightest sources of the full catalogue. However, even if these crude sensitivities help us estimate which source will be seen by ECLAIRs, it can not replace a complete simulation of all the sources through ECLAIRs (see Sec.~\ref{sec:typical_snr}). Such simulation is required to define precisely the list of sources which should be systematically taken into account in the cleaning process and/or blacklisted as known sources for the trigger, should they enter in ECLAIRs' field of view during a given observation.

\subsection{Known sources influence in the detector images before deconvolution}
\label{sec:simu_source}

\begin{figure*}
\centering
\includegraphics[width=17cm]{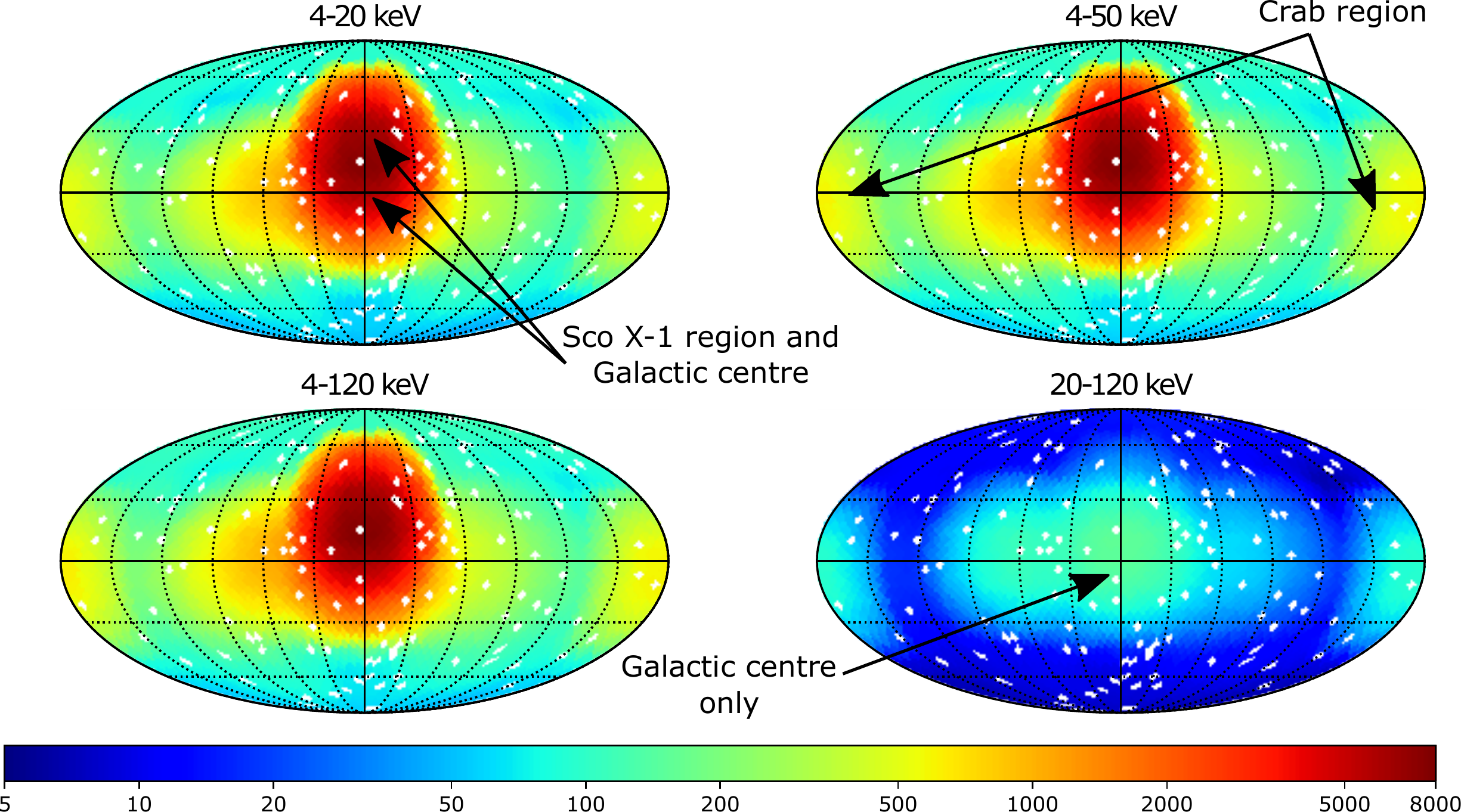}
\caption{Number of counts per second from all the 1793 sources according to the pointing direction in the sky in Galactic coordinates (longitude increasing from right to left). White pixels correspond to regions where no position has been drawn. The four maps correspond to the four energy bands currently defined for the onboard trigger algorithm.}
\label{fig:mapSrcCounts}
\end{figure*}

In order to study the influence of the known X-ray sources on ECLAIRs' noise level, we draw 10000 isotropic pointing positions in the sky (the satellite roll angle is fixed to 0$\degree$). For each of these pointings, we select the list of sources in the field of view from our full catalogue. Each source is projected by ray-tracing from its local coordinates in ECLAIRs' frame. For these simulations, we assume that the sources are persistent at the level of their median flux, even if it is well known that some sources exhibit a variable or bursty behaviour (see Fig.~\ref{fig:lc_src}). The detailed way the sources are processed and cleaned by the onboard software is described in Sec.~\ref{sec:management_cleaning}. For each shadowgram (each exposed for 20 min) we compute the count rate per second added by the sources in the given field of view. The maps giving this rate are shown in Fig.~\ref{fig:mapSrcCounts} for the four different energy bands. In these maps, we indicate some sky regions that will be discussed more in detail in the following (see Fig.~\ref{fig:map} for precise source location). The central zone where the highest rates are reached corresponds to the Galactic centre and the Sco X-1 region.

In Fig.~\ref{fig:mapSrcCounts}, we first notice that the count rate added by the bright sources is quite variable over the whole sky, and roughly goes from $\approx 100$ counts/s close to the Galactic poles to a maximum of $\approx 5000 $ counts/s close to the Galactic center and Sco X-1 (4--120 keV). As a comparison, the CXB contributes to the background level for $\approx 3000$ counts/s in 4--120 keV. The sources contribution is dominant in the first three bands, i.e. 4--20, 4--50, and 4--120 keV, while in the last band, above 20 keV it is much less important. Due to its broken powerlaw spectrum and its high energy photon index of -5.8 above $\approx 10$ keV, the Sco X-1 contribution fades away in the 20--120 keV band. Table~\ref{tab:srcCounts} gives the mean number of counts and its standard deviation for the different energy bands considered and different zones on the sky. The ratio between the mean rate for pointings within the B1 law and outside the B1 law is $\approx 11.1$ for the first three bands and $\approx 4.3$ for the 20--120 keV band. Inside the B1 law, the rate added from the sources remains negligible compared to the CXB, whereas near the Galactic centre and close to Sco X-1, the sources contribution is dominant ($\approx$ 8000 counts is the Galactic centre). Outside the B1 law, the rate is subject to a large dispersion due to the very different regions within the Galactic plane. The centre is much more crowded with bright sources than the other regions.

\begin{table}[!h]
\caption{Mean and standard deviation of the count-rate contributions (in counts/s on ECLAIRs' detection plane) of the sources for the four energy bands considered and for different zones in the sky.}
\label{tab:srcCounts}
\centering
\begin{tabular}{|c|c|c|c|c|}
\hline
\multirow{2}{*}{Energy range} & \multicolumn{2}{c|}{Inside B1} & \multicolumn{2}{c|}{Outside B1} \\ \cline{2-5}
& Mean  & Std & Mean & Std \\
\hline\hline
4--20 keV & 77.2 & 17.3 & 899.1 & 1534.1 \\
4--50 keV & 85.5 & 19.0 & 937.6 & 1557.6 \\
4--120 keV & 88.0 & 19.6 & 946.1 & 1561.0 \\
20--120 keV & 10.8 & 2.4 & 47.0 & 36.0 \\
\hline
\end{tabular}
\end{table}

\subsection{Influence of known sources on the sky images after deconvolution}
\label{sec:sky_images}

To estimate the effect of bright sources on the deconvolved sky images, we used the same 10000 pointing positions in the sky as the ones used in the previous section. For each of these positions, we generate 64 shadowgrams of 20.48 s each, built from the contribution of all our catalogue sources in the field of view, with added CXB and internal background contributions. The 64 shadowgrams are deconvolved and the 64 sky images (each with exposure 20.48 s) are summed up to build all the timescales ($20.48 \times 2^{n-1}$ s with $n=1..7$) to reach a maximum timescale of $\sim 20$ min (1310.72 s), the same way it will be done on board by the image trigger.

We first do not clean the shadowgram for any noise including the CXB. Ideally for a perfectly cleaned shadowgram (or the shadowgram of an absolutely empty sky region), the distribution of the SNR should follow a normal distribution ${\cal{N}}(0,1)$ with zero mean and standard deviation equal to 1. In such sky images, the onboard trigger algorithm can detect new sources (such as GRBs) when the SNR exceeds a threshold of 6.5, with a low false-trigger rate (less than one per day, \citealt{schanne_presentation_2009}). 

In the case where the shadowgram contains uncleaned source patterns before deconvolution, the resulting sky image will contain point sources with high SNR and additional coding noise anywhere in the reconstructed sky where it can not be suppressed anymore (because located away from the positions of these sources). The point sources and the coding noise levels can reach the detection threshold, resulting in false alerts. Even if the source position can be ignored during the image analysis, the coding noise widens the SNR distribution over the whole reconstructed sky, which can lead to false triggers. To avoid this, the trigger threshold would have to be raised. Therefore, if not suppressed prior to deconvolution, the coding noise will reduce the GRB detection efficiency. Also, the standard deviation of the SNR in the reconstructed image (excluding the positions of point sources) is a good indicator of the presence of coding noise in this sky image. Its width needs to remain as close to 1 as possible.

\begin{figure*}
\centering
\includegraphics[width=15cm]{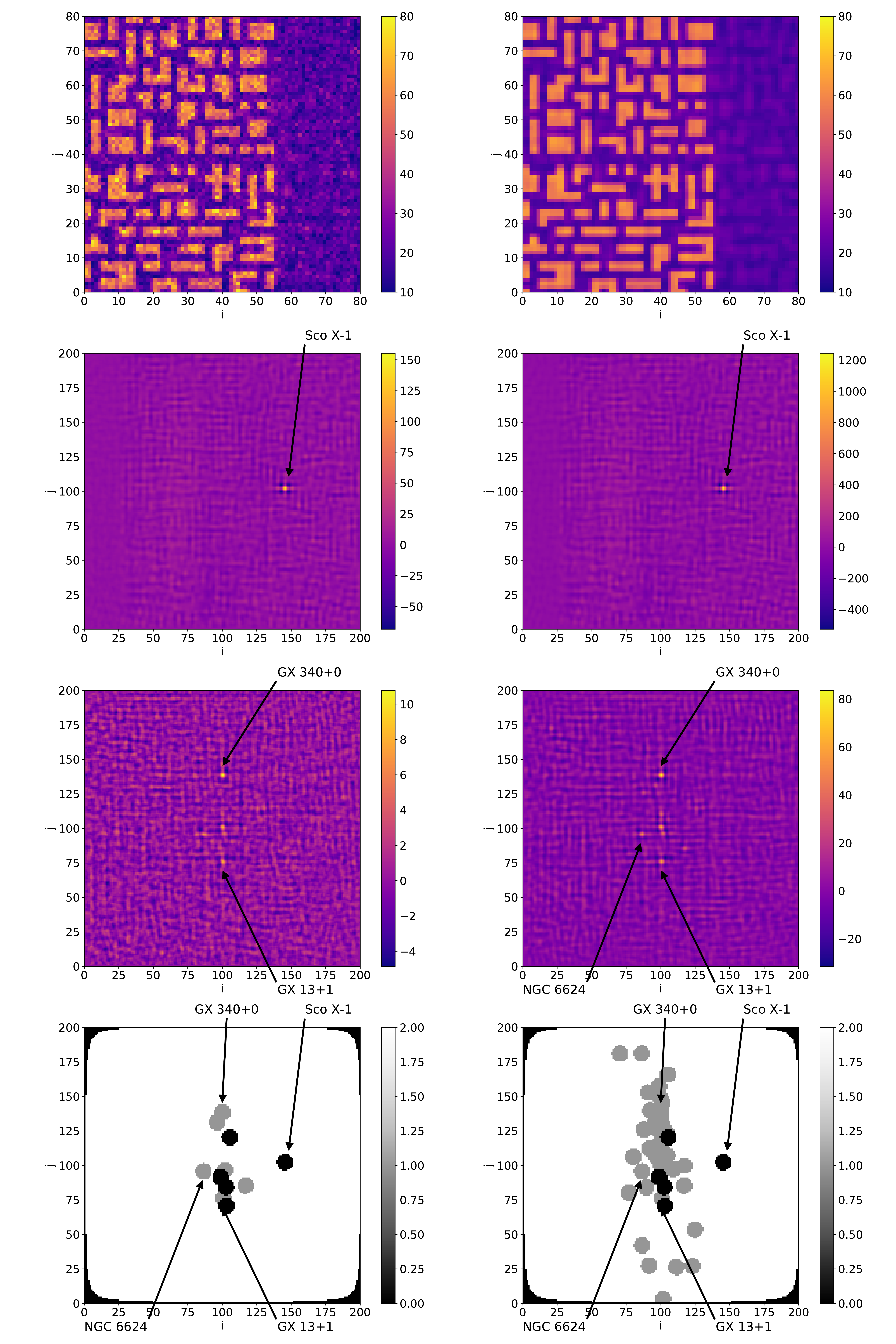}
\caption{First row, left: shadowgram exposed for 20.48 s in the direction of the Galactic centre (4--120 keV). First row, right: corresponding weighted model used for cleaning. Second row: corresponding sky image in SNR in 20 s (left) and in 20 min obtained by the stacking of 64 different 20 s sky images. (right). Third row: same but the shadowgrams are cleaned prior to deconvolution. Last row: sources represented by black (resp. grey) circles are cleaned (resp. only masked in the sky image) in 20 s sky image (left) and 20 min sky image (right).}
\label{fig:det_sky_demo}
\end{figure*}

Figure \ref{fig:det_sky_demo} shows an example of a shadowgram exposed for 20.48 s in the Galactic centre (first row, left) and the associated SNR sky image (second row, left). In the shadowgram, the mask pattern drawn by the bright source Sco X-1 is clearly visible. This pattern leads to a bright point source in the sky image and an SNR distribution (outside of this point source) with a standard deviation as large as 6.06. The other bright sources in the field of view are completely drowned out by this coding noise. An image of $\sim$ 20 min is built by the stacking of 64 contiguous 20 s long sky images and is shown in the same figure (second row, right). Its SNR distribution outside Sco X-1 has a standard deviation of $47.45$.

\begin{figure*}
\centering
\includegraphics[width=17cm]{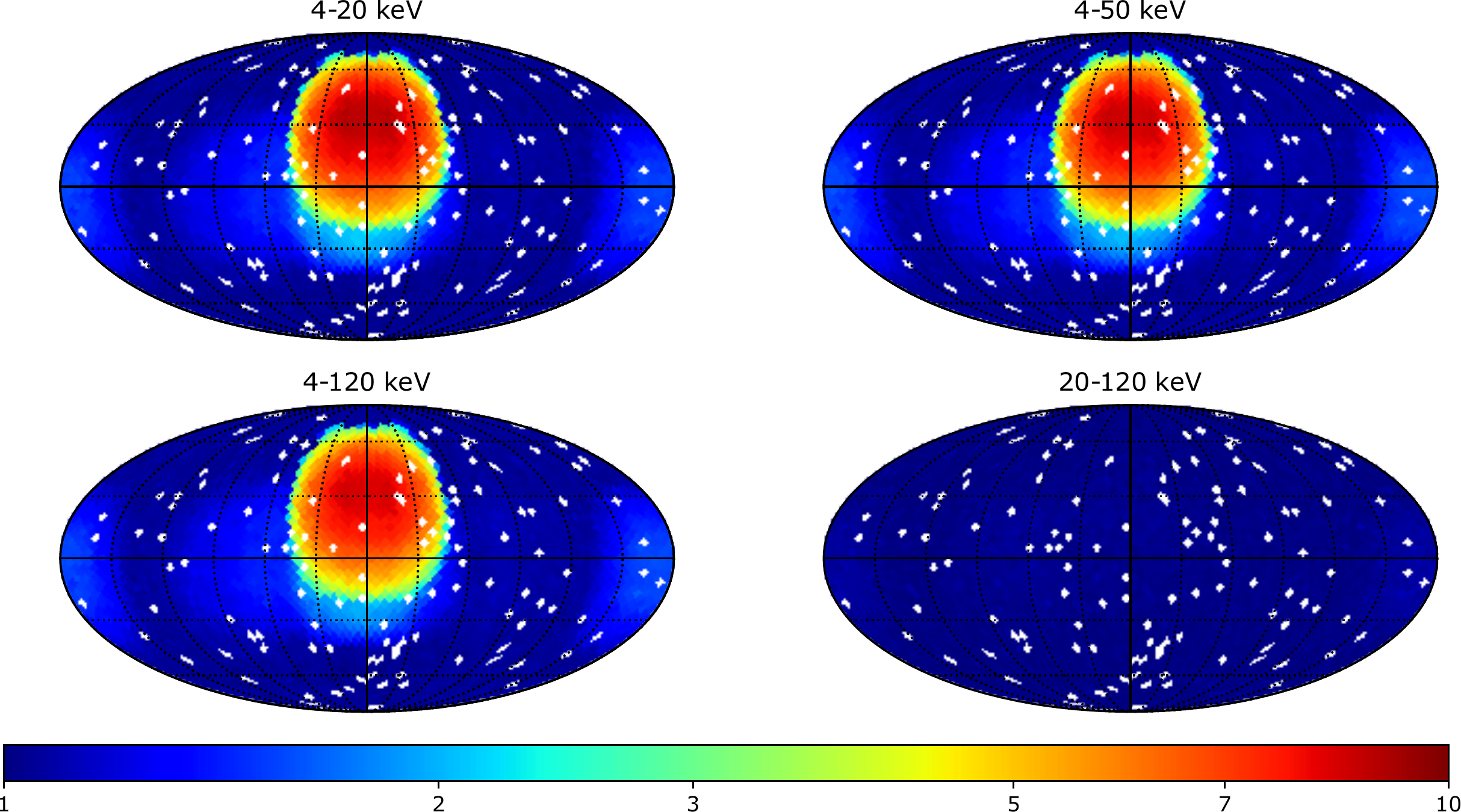}
\caption{Standard deviation of SNR values in 20 s sky images according to the pointing direction in the sky in Galactic coordinates (longitude increasing from right to left). White pixels correspond to regions where no position has been drawn. The four maps correspond to the energy bands foreseen for the onboard trigger algorithm.}
\label{fig:mapStd20s}
\end{figure*}

\begin{figure*}
\centering
\includegraphics[width=17cm]{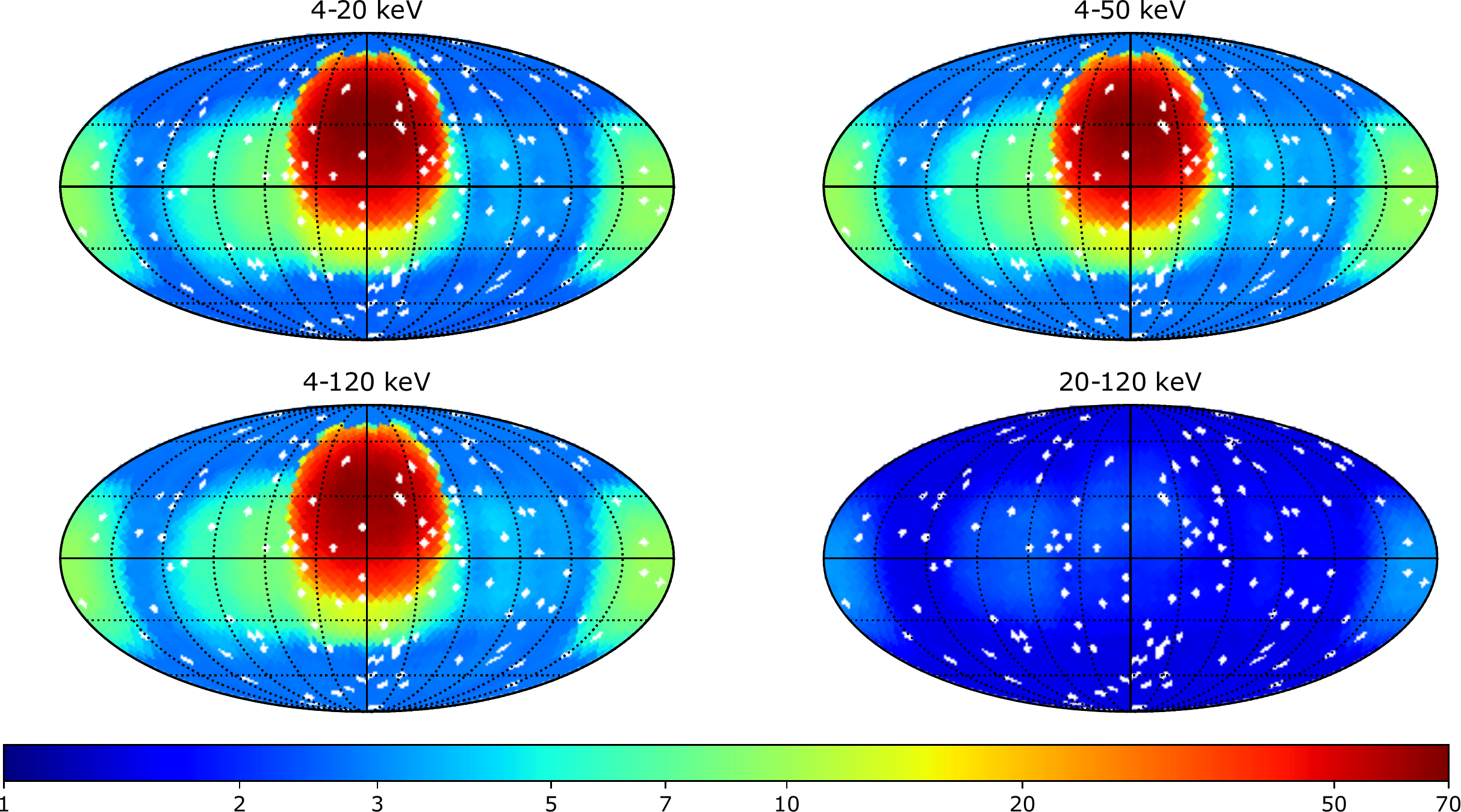}
\caption{Standard deviation of SNR values in 20 min sky images according to the pointing direction in the sky in Galactic coordinates (longitude increasing from right to left). White pixels correspond to regions where no position has been drawn. The four maps correspond to the energy bands foreseen for the onboard trigger algorithm.}
\label{fig:mapStd20min}
\end{figure*}

Figures \ref{fig:mapStd20s} and \ref{fig:mapStd20min} show the SNR standard deviation according to the pointing position in Galactic coordinates for 20~s and 20~min sky images. In these sky images, an unsubtracted CXB background shape before deconvolution also contributes to widen a little bit the SNR distribution: in 20 s exposure images the unsubtracted CXB shape leads to a deviation of $\sim 1$, hence no cleaning of the CXB is required for shorter scales, while the standard deviation increases with the exposure time and reaches $\sim$ 2.3 in the longest scale of 20 min (4--120 keV).

In the 20 s sky images (Fig.~\ref{fig:mapStd20s}), within the B1 law the standard deviation of the SNR is close to 1 in the four bands. Such a deviation is suitable for new source detections in sky images with weak coding noise and an SNR threshold of 6.5. However, in the Galactic centre and near Sco X-1, the standard deviation of the SNR reaches $\sim 10$ in the energy bands starting at 4 keV and is no longer compatible with a detection threshold for new sources of 6.5 $\sigma$. In the 20--120 keV energy band, in 20 s, the standard deviation is close to 1 for each sky pointing direction (the maximum reaches $\approx 1.1$ in the Galactic centre).

In the 20 min sky images (Fig.~\ref{fig:mapStd20min}), within the B1 law the mean standard deviation of the SNR is $\approx 2.8$ in the three first bands and $\approx 1.5$ in the 20--120 keV band. In the Galactic centre and near Sco X-1 the standard deviation of the SNR reaches $\sim 70$ in the energy bands starting at 4 keV and $\sim 2$ in the 20--120 keV band. In any case, such a high SNR standard deviation is not compatible with a detection threshold for new sources of 6.5 $\sigma$. In these sky regions, the detection threshold would have to be raised to much higher levels.

\begin{figure*}
\centering
\includegraphics[width=17cm]{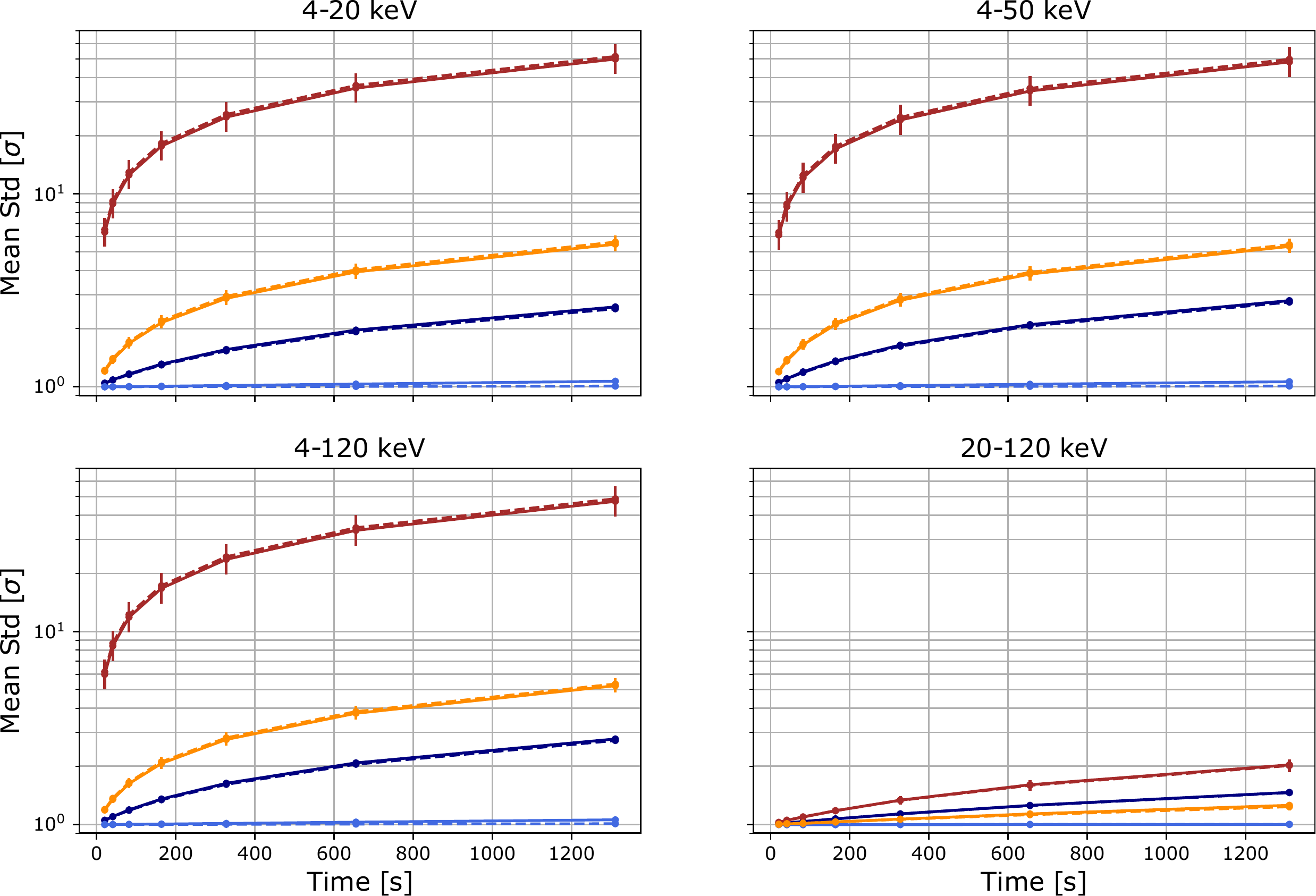}
\caption{Mean standard deviation of SNR values in sky images from different regions in the sky. Blue: within the B1 law without cleaning (dark) and after cleaning (light, see Sec.~\ref{sec:management_cleaning}); within the Galactic centre (box of $\pm15\degree$ in longitude and latitude) without cleaning (red) and after cleaning (orange, see Sec.~\ref{sec:management_cleaning}); solid: simulation of the full catalogue; dotted: simulation of the 53 bright sources only. The four plots correspond to energy bands foreseen for the onboard trigger algorithm.}
\label{fig:meanStdVSTime}
\end{figure*}

Figure \ref{fig:meanStdVSTime} shows the mean standard deviation of the SNR for different regions of the sky: within the B1 law in dark blue or in the Galactic centre (box of $\pm15\degree$  in longitude and latitude) in red. The solid curves (resp. dotted curves) give the result for the simulation of the full catalogue (resp. of the 53 brightest sources). In the four bands, the difference between the two simulations is very small (solid and dotted curves overlap). Within the B1 law (dark blue) the relative difference is $<2\%$. In the Galactic centre the relative difference also remains small ($\approx 3\%$ in the three first bands). Therefore, the contribution of the faint sources remains negligible in comparison to the bright sources. Thus, the reduced catalogue, containing the 53 brightest sources, well describes the influence of the sources on ECLAIRs' background level and on the standard deviation of the SNR in the sky images. However, a simulation of the full catalogue is required to study the detectability of each source (regardless of the sensitivity).

\subsection{Source typical SNR}
\label{sec:typical_snr}

\begin{figure*}
\centering
\includegraphics[width=17cm]{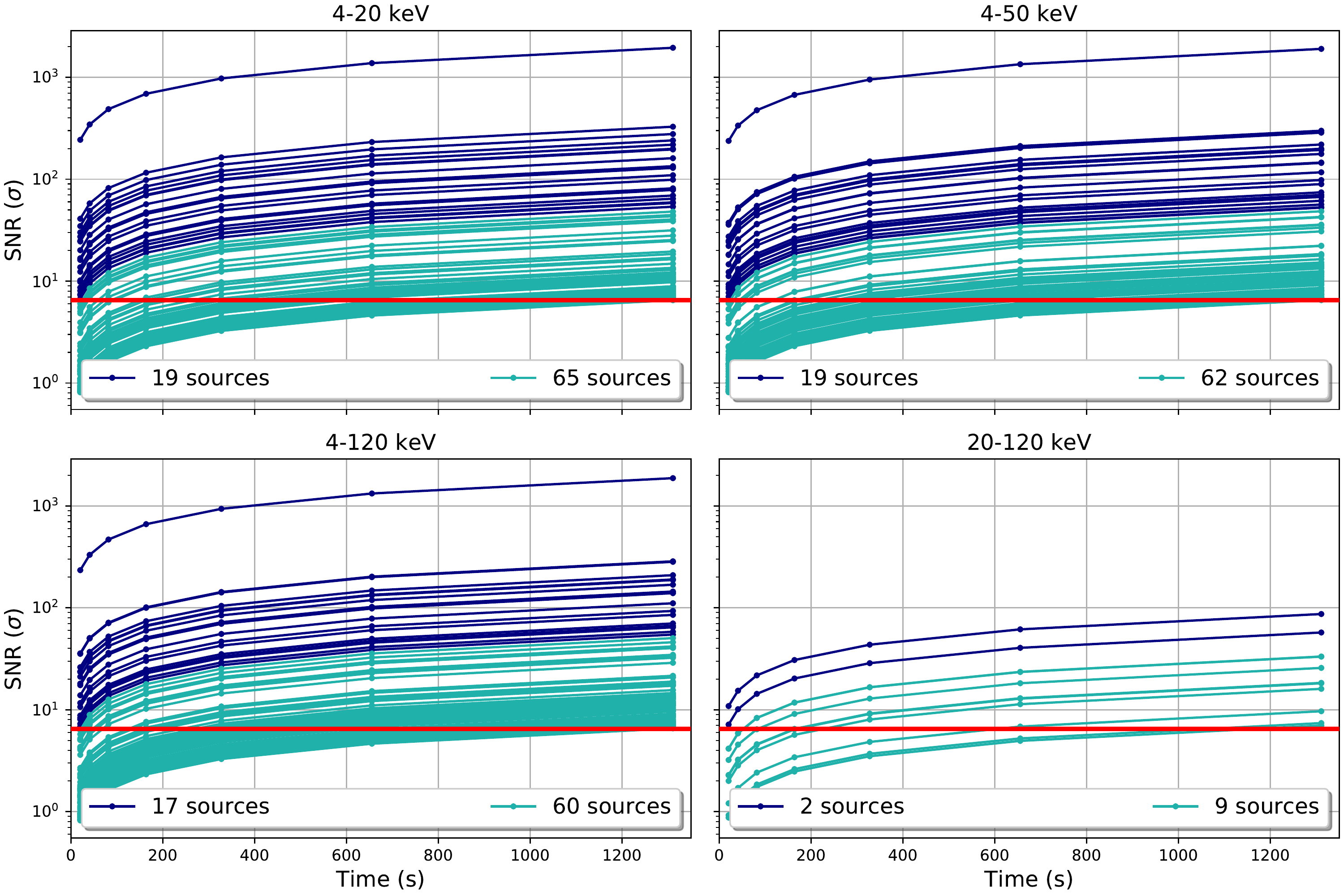}
\caption{Typical SNR of the catalogue sources obtained after simulation of each source in the centre of the field of view. Dark blue (resp. cyan) curves correspond to the sources with SNR $>$ 6.5 in 20 s (resp. 20 min). The red line corresponds to the threshold of 6.5 $\sigma$. The four plots correspond to the energy bands foreseen for the onboard trigger algorithm.}
\label{fig:SNR_sources_1by1}
\end{figure*}

We define the ``typical SNR'' of a source to be the SNR obtained from its simulation using the source spectrum previously determined (see Sec. \ref{sec:spectra}) in the central pixel of the field of view of ECLAIRs with only the CXB and without any other source present. This typical SNR is the expected upper limit of the source's SNR (except in case of source outburst), since it is determined in a very favourable situation: in the part of the field of view with the best sensitivity and without the possible perturbations from other sources. The typical SNR is given in Fig.~\ref{fig:SNR_sources_1by1} for all the sources with SNR$>$ 6.5 in 20 min and for the four energy bands foreseen. The high significance of the points comes from the fact that the sources are simulated in the centre of the field of view. An error of 15$\%$ of the SNR is to be considered when the source position varies within the fully-coded field of view for all energy bands and timescales. This dispersion results from the difference of sensitivity within a sky pixel (the best sensitivity is achieved when the source is in the centre of a sky pixel).

In the four bands, there are respectively 84, 81, 77 and 11 sources with a typical SNR larger than 6.5 in 20 min. In total, 89 sources have a typical SNR larger than 6.5 in 20 min in at least one of the four bands. For these sources, the typical SNR is given in Table~\ref{tab:srclist}. The number of sources in this table is larger than the estimated number of bright sources according to the sensitivity (53). Indeed, the sensitivity was determined for a specific spectrum (simple powerlaw of index $-2$) with a fit of many points in the fluence-duration plane (see Fig.~\ref{fig:sensitivity}). Also, to compute the sensitivity, the dummy sources were not necessarily simulated from the centre of a pixel of the fully-coded field of view. Thus it is possible that a source (potentially with a different spectrum) produces an SNR larger than 6.5 even if its flux is slightly below the sensitivity limit.

\subsection{Source management and cleaning}
\label{sec:management_cleaning}

As we showed in the previous sections, the known sources will have an important impact on the background level, the sky image quality and therefore the capability to detect new sources above a threshold of 6.5 $\sigma$ with a low false trigger rate \citep{schanne_presentation_2009}. An onboard cleaning step is therefore required during the imaging process. It is beyond the scope of this paper to precisely describe the methods to clean the source contributions and their performances. However, we provide in this section an example of what can be expected after cleaning. In order to demonstrate that the quality of the reconstructed sky images benefits from the cleaning of the sources, we carry out the same simulations as in Sec.~\ref{sec:sky_images} to which we add a cleaning step before the deconvolution. The shadowgrams are cleaned by fitting and subtracting a model composed of a quadratic-shape model of the CXB plus the illumination functions of up to 5 sources present in the field of view. The number of sources which will be modelled is limited to 5 because of limited computation resources on board, and to reduce the risk of divergence of the fit. For the sources to be modelled, the onboard algorithm selects, after each repointing, the 5 strongest sources in the field of view and computes their illumination function (projected shadowgram model) used in the fit. By strongest sources we mean the sources that project the largest number of photons taking into account the source flux but also the source position in the field of view (a bright source in a corner of the field of view may project less photons than a fainter source in the centre of the field of view).

Figure \ref{fig:det_sky_demo} (first row, right), gives an example of a weighed shadowgram model obtained after the fit of the raw shadowgram. The weighted model is then subtracted from the raw shadowgram to get a cleaned shadowgram ready for deconvolution. Figure \ref{fig:det_sky_demo} (third row, left), gives the sky image in SNR after deconvolution of the cleaned shadowgram. This image is to be compared with the one shown on the same figure (second row, left) where no cleaning has been performed prior to the deconvolution. The standard deviation of the SNR in the sky region excluding the point sources is 1.14 (compared to 6.06 in the sky image from the raw shadowgram shown in the second row, left). In this shadowgram, the 5 sources that are cleaned are the low-mass X-ray binaries Sco X-1, GX 5-1, GX 349+2, GX 9+1 and GX 17+2 (represented by black circles in the last row of Fig.~\ref{fig:det_sky_demo}; the Galactic plane is vertical). After cleaning and deconvolution, some sources that where hidden by the Sco X-1 coding noise have now become visible (GX 340+0, GX13+1). The stacking up to 20 min reveals some fainter sources that are not cleaned in the shadowgram (Sgr X-4).

\begin{figure*}
\centering
\includegraphics[width=17cm]{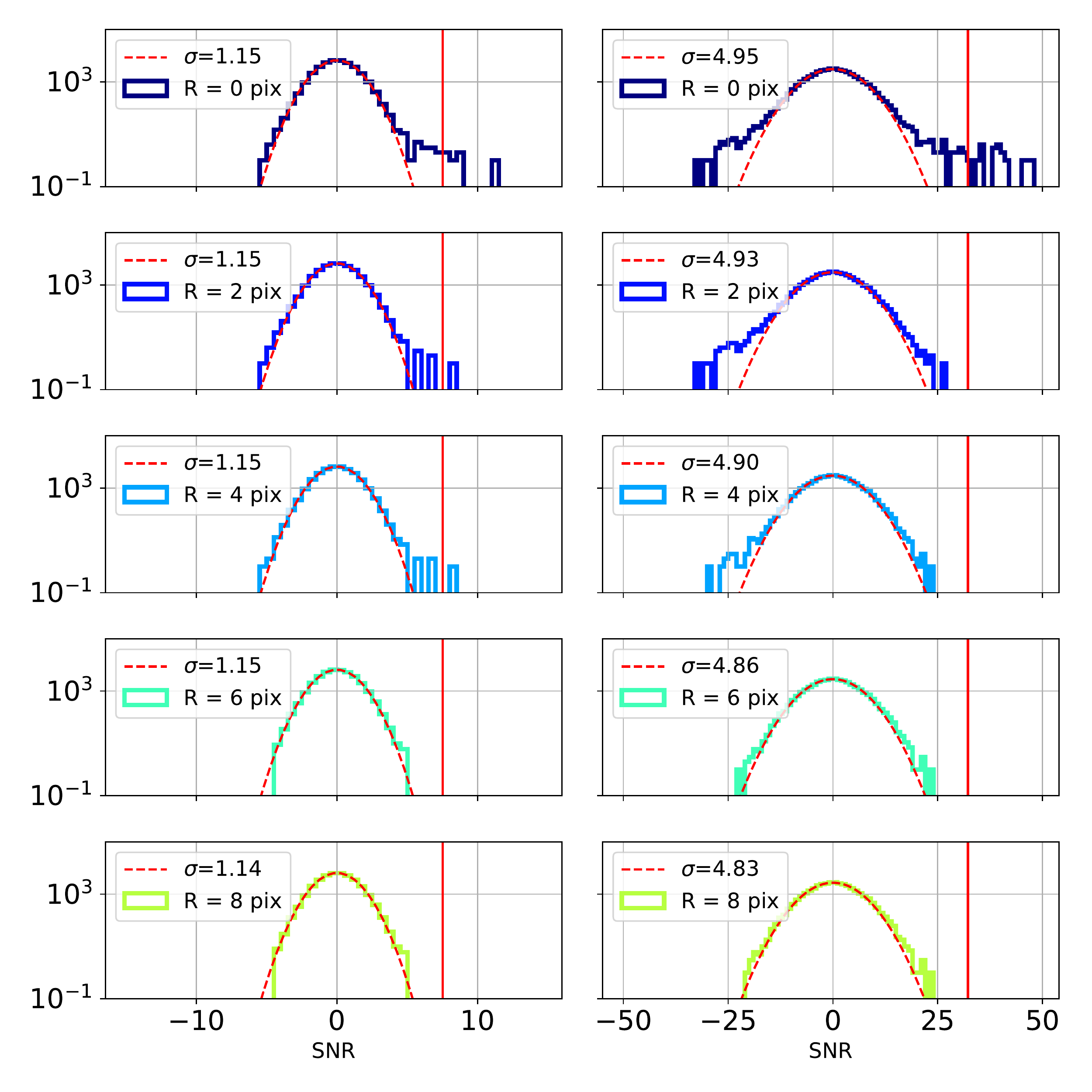}
\caption{Distributions of the SNR in a sky image after deconvolution of a cleaned shadowgram for different radii used to exclude the sources. Left column: distribution of the SNR in the 20 s sky image of Fig.~\ref{fig:det_sky_demo} (centre). Right column: distribution in 20 min sky image (same Fig., right). The red dashed lines correspond to a Gaussian function fitted to the distribution. With these distributions, the threshold to be applied for source detection would be 1.15$\times$6.5 = 7.5 for the 20 s case, and 4.86$\times$6.5 = 31.6 for the 20 min case, as indicated by the red vertical lines.}
\label{fig:histo_img_20s_min_R}
\end{figure*}

To get more details on the quality of the deconvolved sky image, which exhibits some point sources, and their coding noise in the region outside of the sources, we study the distribution of the SNR of the image pixels. Figure \ref{fig:histo_img_20s_min_R} shows the histograms of the sky images shown in Fig.~\ref{fig:det_sky_demo} (left column for 20 s and right column for 20 min) for different radii used to exclude the sources. If the point sources are not excluded (first row), the distribution is characterised by a Gaussian part plus tails of extreme SNR values which correspond to the point sources. The tails can be removed by excluding enough pixels around the source positions. The optimal radius to exclude and efficiently remove the tails is 6 sky pixels, which suits both for the 20 s and 20 min sky images. With smaller radii, residual high SNR values too close to the strong sources may still cause false detections. Moreover, for a sky image resulting from a cleaned shadowgram deconvolution, when the point sources are excluded with radii sufficiently large to remove the tails, the distribution of the SNR values in the image is well approximated by a Gaussian distribution of mean zero and standard deviation $\sigma$. For a Gaussian distribution, the measured standard deviation in the image is equal to the $\sigma$ of the Gaussian. Its benefit is then to be able to configure the detection threshold as a function of this $\sigma$ to get an acceptable false alert rate.

Our analysis method will be applied on board by the trigger algorithm to be able to operate in sky regions with strong sources: after sky deconvolution, it will determine the standard deviation $\sigma$ of the distribution of the SNR in the sky pixels excluding the sources, and adapt the trigger threshold to be actually above the coding noise with a safety factor (e.g. 6.5$\times \sigma$).

\begin{figure*}
\centering
\includegraphics[width=17cm]{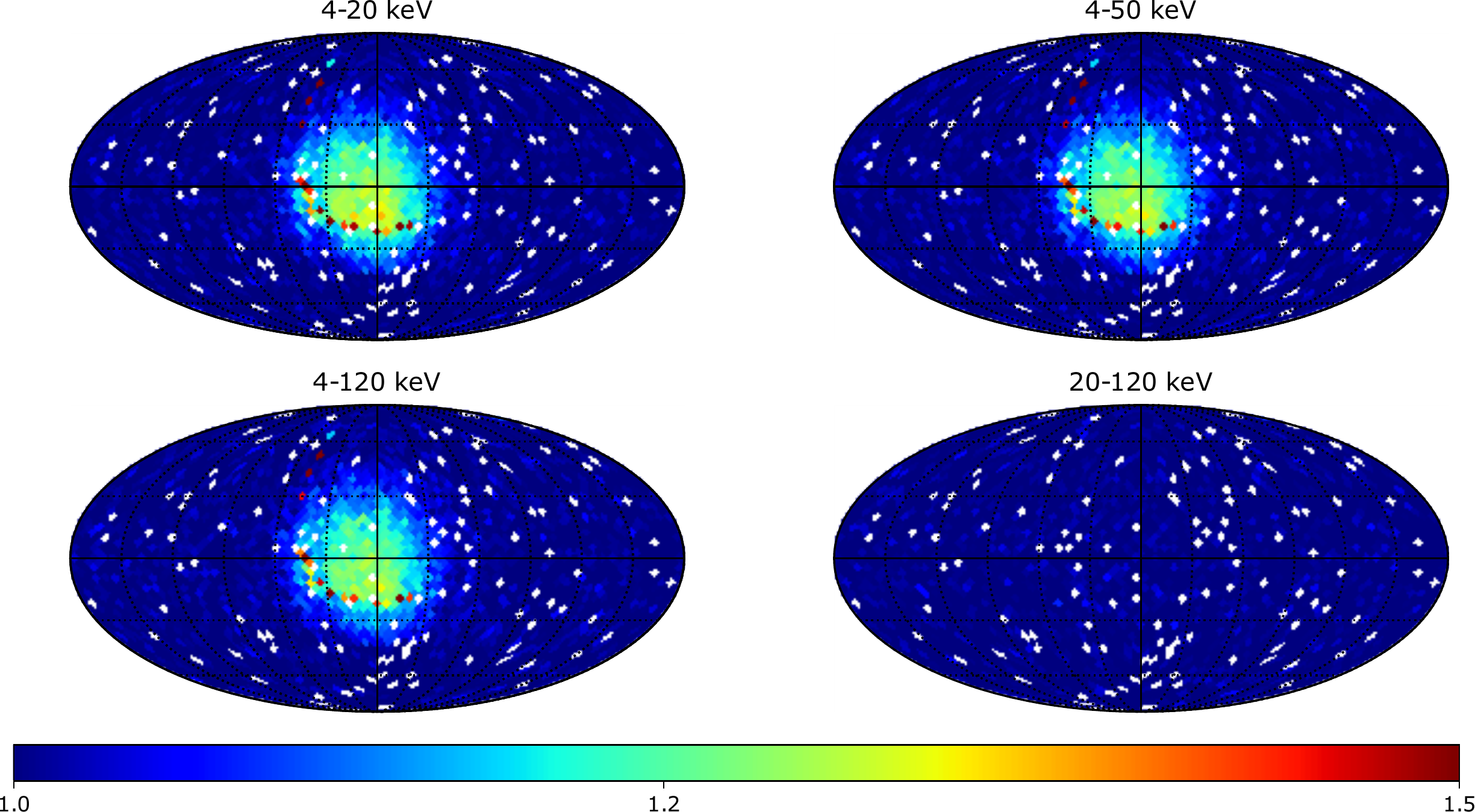}
\caption{Standard deviation of SNR values in 20 s sky images (after the shadowgram cleaning) according to the pointing direction in the sky in Galactic coordinates (longitude increasing from right to left). White pixels correspond to parts where no position has been drawn. The four maps correspond to energy bands foreseen to be defined for the onboard trigger algorithm.}
\label{fig:mapSrcStd20sClean}
\end{figure*}

Figure \ref{fig:mapSrcStd20sClean} shows the standard deviation of the SNR according to the pointing position in Galactic coordinates for 20 s sky images, where the shadowgrams have been cleaned prior to deconvolution the way we described previously. In the Galactic centre, in the three first energy bands, the standard deviation of the SNR reaches $\approx 1.2$ (compared to $\approx 6$ without cleaning) whereas in the other regions it stays close to 1. In the 20--120 keV energy band, it remains close to 1 for all directions. The artefacts that can be seen near the Galactic centre result from the non cleaning of Sco X-1 when this source is close to a border of the field of view and does not belong to the 5 sources to clean. This effect illustrates the limitations of the strategy based only on the fit to clean the sources. To prevent this issue which arises when Sco X-1 is not classified among the 5 strongest sources to be fitted, the detector pixels which are illuminated by Sco X-1 can be discarded from the fit procedure and also from the deconvolution by attributing them a weight of zero. When Sco X-1 is very partially coded, this method to suppress Sco X-1 does not reduce much the number of pixels active in the imaging process, and thus does not degrade much the sensitivity.

\begin{figure*}
\centering
\includegraphics[width=17cm]{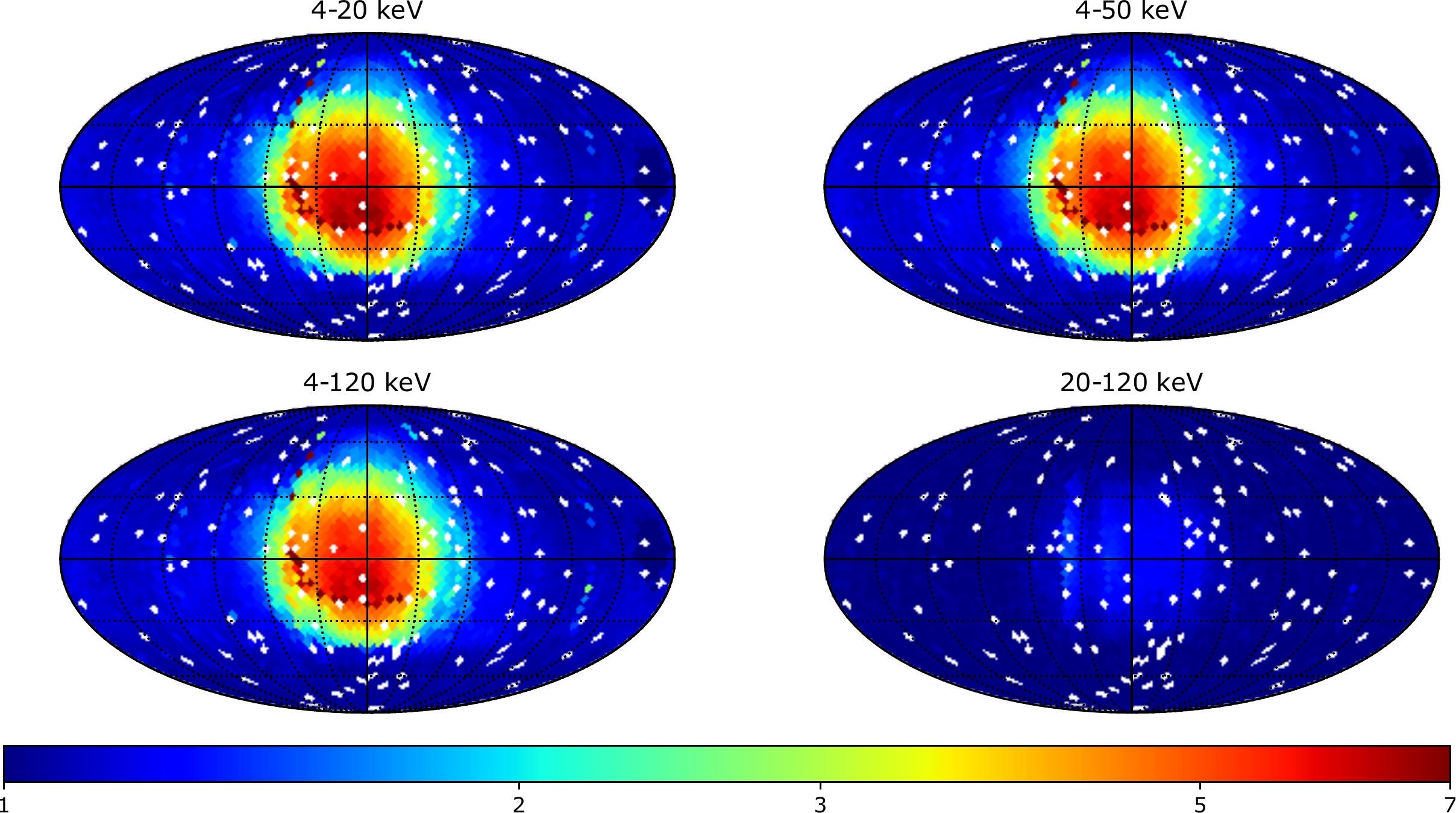}
\caption{Standard deviation of SNR values in 20 min sky images (after the shadowgram cleaning) according to the pointing direction in the sky in Galactic coordinates (longitude increasing from right to left). White pixels correspond to parts where no position has been drawn. The four maps correspond to energy bands foreseen to be defined for the onboard trigger algorithm.}
\label{fig:mapSrcStd20minClean}
\end{figure*}

Figure \ref{fig:mapSrcStd20minClean} shows the standard deviation of the SNR according to the pointing position in Galactic coordinates for 20 min sky images, where the shadowgrams have been cleaned prior to deconvolution the way we described previously. Figure \ref{fig:meanStdVSTime} also shows the evolution of the mean value of the standard deviation as a function of the timescale after cleaning (orange curves for the Galactic centre and light blue within the B1 law). Thanks to the cleaning, the dispersion of the SNR for cleaned images of 20 min duration approaches the one obtained for images of 20 s without cleaning. Within the B1 law the standard deviation of the SNR is $\approx 1.1$ in the three first bands and $\approx 1$ in the 20--120 keV band. This is well adapted to the detection of sources above a threshold of 6.5 $\sigma$. However, in the Galactic centre region, the standard deviation still reaches $\approx 5.7$ in the three first bands and $\approx 1.3$ in the last band. This sky region would require further processing if possible, or needs the usage of a higher SNR threshold to search for new sources with an acceptable false detection rate.

We would like to mention that the method we briefly describe here is not the only possible one. Indeed, the CXB can also be cleaned using a wavelet filtering of the shadowgrams. Moreover, a source contribution can be completely removed by attributing in the deconvolution step a weight of zero to all the detector pixels illuminated by this source (this is particularly important for Sco X-1 when it has a small coding fraction). The performances of these alternative methods will be discussed in a future paper.

\section{Onboard catalogue}
\label{sec:structure}

\subsection{Catalogue structure}

The details of the onboard catalogue of sources for ECLAIRs are still under development and are beyond the scope of this paper. In this section we give an overview of the possible information that will be provided in the detailed catalogue. 

The catalogue table contains the sources that are bright enough to be detected by ECLAIRs in the maximum timescale of 20 min and deserve some processing. A flag will be given to set the processing for each source. The flag \textit{det} is set for the brightest sources that can be detected in 20 s sky images and will be subtracted from the shadowgram using the fit method; alternatively, the detector pixels they illuminate can be discarded for the deconvolution step. The flag \textit{sky} is set for the sources that can be detected in the stacked sky images up to 20 min and will be ignored in the sky image during the new sources search. In the case where one of the sources bearing the flag \textit{sky} in the catalogue is in a flaring state, the triggering system will be able to detect it above an SNR threshold also given in the catalogue.

The flag which determines the source processing is attributed to each source according to its typical SNR (see Sec.~\ref{sec:typical_snr}): in 20 s for the \textit{det} flag and in 20 min for the \textit{sky} flag. Table \ref{tab:srclist} gives the flags in the different bands for the brightest sources (an absence of flag in one of the energy bands means that the source will not be cleaned or avoided in this band and thus can be detected as a new source).

\begin{figure*}
\centering
\includegraphics[width=17cm]{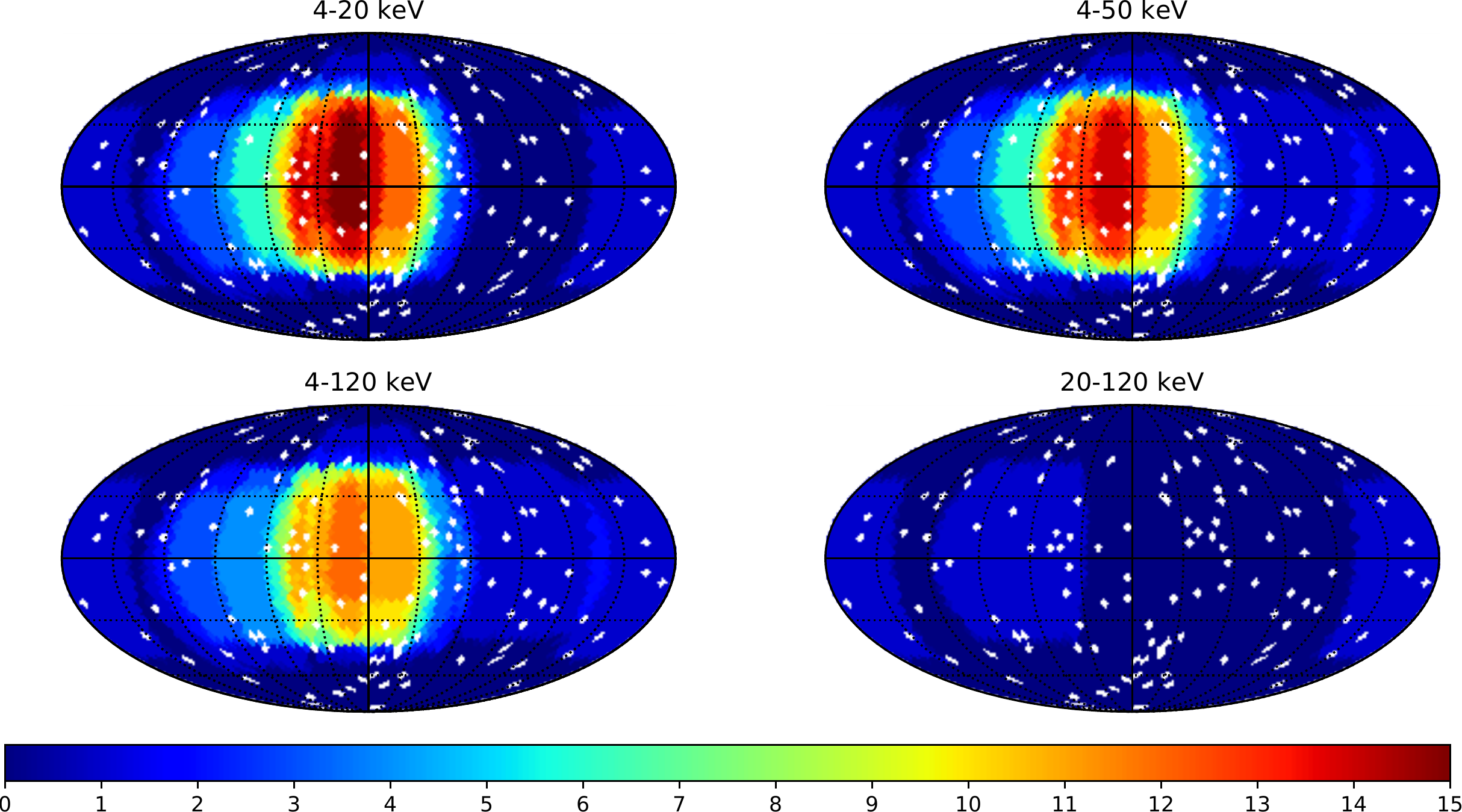}
\caption{Number of sources, present in the ECLAIRs field of view and bearing the flag \textit{det}, according to the pointing direction in the sky in Galactic coordinates (longitude increasing from right to left). White pixels correspond to part where no position has been drawn. The four maps correspond to the energy bands foreseen for the onboard trigger algorithm. In order to compute the number of sources per sky region, 10000 pointing positions have been drawn isotropically in the sky, with a roll angle fixed to 0$\degree$.}
\label{fig:NtoFit}
\end{figure*}

Figure \ref{fig:NtoFit} shows the number of sources given in Table~\ref{tab:srclist} and bearing the flag \textit{det}, which are present in ECLAIRs' field of view, as a function of the pointing direction in Galactic coordinates. The Galactic centre region contains most of these sources. The region of the sky in which more than 5 sources deserve to be fitted covers respectively 2.68, 2.62 and 2.21 sr in the first three energy bands (which is approximately the size of the total field of view of ECLAIRs, i.e. $1/6$th of the sky). In the last band, only Crab and Cyg X-1 bear the flag \textit{det}. In the case where more than 5 sources should be fitted, the onboard algorithm has to compute a score to determine the 5 best sources to be included in the fit (as discussed previously, see Sec.~\ref{sec:management_cleaning}).

\subsection{Onboard source detection}
\label{sec:src_detection}

The SNR threshold used to detect a source outburst (for the sources bearing the flag \textit{sky} in the onboard catalogue) is based on the source's typical SNR (that can be multiplied by an adjustable factor $K$). Ideally, the threshold should be different for each source, for each energy band and for each timescale, which leads to many parameters in the catalogue. A workaround to reduce the size of the catalogue is to set an SNR threshold for each source, for each energy band and for a reference timescale of 20 min and to compute on board the threshold for the other timescales by applying a scaling factor $S$. The scaling factor can be the same for the four energy bands and for all the sources. It is derived from the SNR evolution according to the exposure time (see Fig.~\ref{fig:SNR_sources_1by1}), normalised by the SNR in 20 min. Thus, for the 7 different timescales ($2^{n-1}\times$ 20.48 s with $n$=1..7), the scaling factor $S$ is : 0.125, 0.177, 0.250, 0.354, 0.500, 0.707, 1.0. As an example, to detect a flare from the source Ser X-1 (in 4--120 keV) whose typical SNR in 20 min is $\approx 50$, the SNR threshold in an image of 20.48 s would be defined as $SNR_{\mathrm{thres}} = K \times 50 \times 0.125 \times \sigma = 6.25 \times K \times \sigma$. In an image from a well cleaned shadowgram ($\sigma = 1$) and for a factor $K=1.5$ (as an example), this leads to a threshold of 9.38 (where the typical SNR is this timescale is 6.25). Once a flare from this source is detected above this threshold, the source's typical SNR can be increased automatically on board to the SNR of the flare, which will provide a larger threshold for a possible future flare, and which prevents triggering multiple times on the source during its outburst which may last many hours or days. 

If an event is detected and its position does not match with a source of the onboard catalogue (or there is no \textit{det} or \textit{sky} flag associated to the source in the triggering energy band), the observation strategy will follow the same as the one defined for GRB candidates. A faint X-ray source, that is not included in the onboard catalogue, may be identified on ground by a search in object databases such as SIMBAD. In this case, the detected source can be added to the onboard catalogue with a flag and a threshold (from ground by telecommand). Here, we suppose that the onboard instruments MXT and VT do not need to know whether the alert comes from such a source or from a GRB.

\subsection{Discussion}

The catalogue that we produced in this study is a preliminary version intended for trigger studies and development. The first onboard operational version of the catalogue will depend on various parameters such as the energy bands which will actually be used by the triggering system, and on the strategy for the detection of outbursts which will be chosen for sources one by one.

In order to build our catalogue, we used for each source its median flux to determine a typical SNR that allowed us to derive a processing flag and a detection threshold. 
However, the sources are not all steady but can exhibit different types of variability (with features such as bright outbursts or periodic flares). 
As a consequence, a few sources of the catalogue presented here have been marked as requiring to be cleaned or avoided, only due to the fact that their lighcurves showed a few bright flares over many years of observation (sufficiently bright to rise their median flux, leading to a typical SNR above 6.5 in 20 min).
Such cases are indicated as ``outbursts'' in the ``class of variability'' attributed to each source in Table~\ref{tab:srclist}, taken from \cite{krimm_swift/bat_2013}.
As an example, the source MAXI J1820+070 is an X-ray binary that entered into a bright outburst phase between March and October 2018 \citep{bharali_broad-band_2019}. It is now in a quiet state but it may produce other outbursts in the future.
These sources will be removed from the operational catalogue as they are not expected to disturb the triggering system. 
Some of these sources are even of particular interest for their rare flares, while being otherwise very faint, and can therefore be treated onboard as unknown sources.

The flags and the thresholds for the detection of known-source outbursts will be set in the onboard catalogue according to the values we computed, but also based on the first data that will be collected by ECLAIRs in flight. 
Indeed all the counts recorded by the detector will be sent to the ground, with a delay of about 6 to 12 hours, which will permit reprocessing of the data and tuning of the onboard algorithms.
During the commissioning phase of the mission (in the first months after launch), it is probable that the autonomous satellite slews towards detections of known-source outbursts will be disabled, and that at the beginning pointings towards the Galactic centre region will be avoided, in order to focus on the tuning of the parameters for the detection of unknown sources (GRB candidates or known sources that are too faint to be in the onboard catalogue).

The onboard catalogue, its number of entries and the values it contains, its flags and thresholds, will be configurable by telecommands from the ground. In particular once the algorithm has automatically  increased the detection threshold for a known source (or included a new unknown source in the catalogue) in order to avoid repeated triggering on the same source outburst, its threshold can be reprogrammed from ground to normal values during the next routine telecommand upload, typically every week.

\section{Conclusion}

The catalogue of sources that we presented in this article is composed of 1793 sources and is built from \textit{Swift}/BAT and \textit{MAXI}/GSC data. We used this catalogue to study the influence of the known X-ray sources on the \textit{SVOM}/ECLAIRs background level and showed that the source contributions will widen the distribution of the SNR in sky images produced after deconvolution of ECLAIRs' detector images. We also produced a preliminary version of the onboard catalogue with 89 bright sources that need to be cleaned or avoided during new-source searches.

The influence of the sources is strongly dependent on the pointing position, on the exposure time and on the trigger energy band. This study shows that in any case, a cleaning of bright sources (such as Sco X-1) is required to reduce as much as possible the coding noise in the sky images and to keep the standard deviation of the SNR distribution in reconstructed sky images close to 1, in order to detect new sources (such as GRBs) as part of the \textit{SVOM} \textit{core program} with a low false trigger rate.

\begin{acknowledgements}
ECLAIRs is a cooperation between CNES, CEA and CNRS, with CNES acting as prime contractor. This work is supported by CEA and by the ``IDI 2017'' project of the French ``Investissements d'Avenir'' program, financed by IDEX Paris-Saclay, ANR-11-IDEX-0003-02. This research has made use of the SIMBAD database, operated at CDS, Strasbourg, France and of the \textit{MAXI} data provided by RIKEN, JAXA and the \textit{MAXI} team.
We would like to acknowledge helpful comments on the manuscript from the anonymous referee. We would also like to thank very much David Palmer from the \textit{Swift}/BAT onboard trigger software team for his very valuable comments and advises.
\end{acknowledgements}

\bibliographystyle{aasjournal}
\bibliography{references}  

\begin{thebibliography}{}
\expandafter\ifx\csname natexlab\endcsname\relax\def\natexlab#1{#1}\fi
\providecommand{\url}[1]{\href{#1}{#1}}
\providecommand{\dodoi}[1]{doi:~\href{http://doi.org/#1}{\nolinkurl{#1}}}
\providecommand{\doeprint}[1]{\href{http://ascl.net/#1}{\nolinkurl{http://ascl.net/#1}}}
\providecommand{\doarXiv}[1]{\href{https://arxiv.org/abs/#1}{\nolinkurl{https://arxiv.org/abs/#1}}}

\bibitem[{Baumgartner {et~al.}(2013)Baumgartner, Tueller, Markwardt, Skinner,
  Barthelmy, Mushotzky, Evans, \& Gehrels}]{baumgartner_70_2013}
Baumgartner, W.~H., Tueller, J., Markwardt, C.~B., {et~al.} 2013, ApJS, 207,
  \dodoi{10.1088/0067-0049/207/2/19}

\bibitem[{Bekhti {et~al.}(2016)Bekhti, Fl{\"o}er, Keller, Kerp, Lenz, Winkel,
  Bailin, Calabretta, Dedes, Ford, Gibson, Haud, Janowiecki, Kalberla, Lockman,
  {McClure-Griffiths}, Murphy, Nakanishi, Pisano, \&
  {Staveley-Smith}}]{bekhti_hi4pi_2016}
Bekhti, N.~B., Fl{\"o}er, L., Keller, R., {et~al.} 2016, A\&A, 594,
  \dodoi{10.1051/0004-6361/201629178}

\bibitem[{Bharali {et~al.}(2019)Bharali, Chauhan, \&
  Boruah}]{bharali_broad-band_2019}
Bharali, P., Chauhan, J., \& Boruah, K. 2019, MNRAS, 487,
  \dodoi{10.1093/mnras/stz1686}

\bibitem[{Connaughton {et~al.}(2015)Connaughton, Briggs, Goldstein, Meegan,
  Paciesas, Preece, {Wilson-Hodge}, Gibby, Greiner, Gruber, Jenke, Kippen,
  Pelassa, Xiong, Yu, Bhat, Burgess, Byrne, Fitzpatrick, Foley, Giles, Guiriec,
  {van der Horst}, {von Kienlin}, McBreen, McGlynn, Tierney, \&
  Zhang}]{connaughton_localization_2015}
Connaughton, V., Briggs, M.~S., Goldstein, A., {et~al.} 2015, ApJS, 216,
  \dodoi{10.1088/0067-0049/216/2/32}

\bibitem[{Dagoneau {et~al.}(2020)Dagoneau, Schanne, Atteia, G{\"o}tz, \&
  Cordier}]{dagoneau_ultra-long_2020}
Dagoneau, N., Schanne, S., Atteia, J.-L., G{\"o}tz, D., \& Cordier, B. 2020,
  Exp Astron, 50, \dodoi{10.1007/s10686-020-09665-w}

\bibitem[{{dal Fiume} {et~al.}(1998){dal Fiume}, Orlandini, Cusumano, {del
  Sordo}, Feroci, Frontera, Oosterbroek, Palazzi, Parmar, Santangelo, \&
  Segreto}]{dal_fiume_broad-band_1998}
{dal Fiume}, D., Orlandini, M., Cusumano, G., {et~al.} 1998, A\&A, 329.
\newblock \url{http://adsabs.harvard.edu/abs/1998A\%26A...329L..41D}

\bibitem[{{Gaia Collaboration}(2018)}]{gaia_collaboration_vizier_2018}
{Gaia Collaboration}. 2018, VizieR Online Data Catalog, 1345.
\newblock \url{http://adsabs.harvard.edu/abs/2018yCat.1345....0G}

\bibitem[{Galloway {et~al.}(2020)Galloway, in~'t Zand, Chenevez, W{\"o}rpel,
  Keek, Ootes, Watts, Gisler, {Sanchez-Fernandez}, \&
  Kuulkers}]{galloway_multi-instrument_2020}
Galloway, D.~K., in~'t Zand, J., Chenevez, J., {et~al.} 2020, ApJS, 249,
  \dodoi{10.3847/1538-4365/ab9f2e}

\bibitem[{Godet {et~al.}(2014)Godet, Nasser, Atteia, Cordier, Mandrou, Barret,
  Triou, Pons, Amoros, Bordon, Gevin, Gonzalez, G{\"o}tz, Gros, Houret,
  Lachaud, Lacombe, Marty, Mercier, Rambaud, Ramon, Rouaix, Schanne, \&
  Waegebaert}]{takahashi_x-gamma-ray_2014}
Godet, O., Nasser, G., Atteia, J.-., {et~al.} 2014, in {{SPIE Astronomical
  Telescopes}} + {{Instrumentation}}, {Montr\'eal, Quebec, Canada},
  \dodoi{10.1117/12.2055507}

\bibitem[{Goldwurm {et~al.}(2003)Goldwurm, David, Foschini, Gros, Laurent,
  Sauvageon, Bird, Lerusse, \& Produit}]{goldwurm_integral/ibis_2003}
Goldwurm, A., David, P., Foschini, L., {et~al.} 2003, A\&A, 411,
  \dodoi{10.1051/0004-6361:20031395}

\bibitem[{Jaubert {et~al.}(2017)Jaubert, Morand, \&
  Jouret}]{jaubert_realistic_2017}
Jaubert, J., Morand, V., \& Jouret, M. 2017, Realistic Mission Scenariosfor
  Satellite Power Analysis, Tech. rep.

\bibitem[{Krimm {et~al.}(2013)Krimm, Holland, Corbet, Pearlman, Romano, Kennea,
  Bloom, Barthelmy, Baumgartner, Cummings, Gehrels, Lien, Markwardt, Palmer,
  Sakamoto, Stamatikos, \& Ukwatta}]{krimm_swift/bat_2013}
Krimm, H.~A., Holland, S.~T., Corbet, R. H.~D., {et~al.} 2013, ApJS, 209,
  \dodoi{10.1088/0067-0049/209/1/14}

\bibitem[{Le~Provost {et~al.}(2013)Le~Provost, Schanne, Flouzat, Kestener,
  Chaminade, Donati, Ch{\^a}teau, Daly, \&
  Fontignie}]{le_provost_scientific_2013}
Le~Provost, H., Schanne, S., Flouzat, C., {et~al.} 2013, in 2013 {{IEEE Nuclear
  Science Symposium}} and {{Medical Imaging Conference}},
  \dodoi{10.1109/NSSMIC.2013.6829557}

\bibitem[{Mate {et~al.}(2019)Mate, Bouchet, Atteia, Claret, Cordier, Dagoneau,
  Godet, Gros, Schanne, \& Triou}]{mate_simulations_2019}
Mate, S., Bouchet, L., Atteia, J.-L., {et~al.} 2019, Exp Astron, 48,
  \dodoi{10.1007/s10686-019-09643-x}

\bibitem[{Matsuoka {et~al.}(2009)Matsuoka, Kawasaki, Ueno, Tomida, Kohama,
  Suzuki, Adachi, Ishikawa, Mihara, Sugizaki, Isobe, Nakagawa, Tsunemi, Miyata,
  Kawai, Kataoka, Morii, Yoshida, Negoro, Nakajima, Ueda, Chujo, Yamaoka,
  Yamazaki, Nakahira, You, Ishiwata, Miyoshi, Eguchi, Hiroi, Katayama, \&
  Ebisawa}]{matsuoka_maxi_2009}
Matsuoka, M., Kawasaki, K., Ueno, S., {et~al.} 2009, PASJ, 61,
  \dodoi{10.1093/pasj/61.5.999}

\bibitem[{Moretti {et~al.}(2009)Moretti, Pagani, Cusumano, Campana, Perri,
  Abbey, Ajello, Beardmore, Burrows, Chincarini, Godet, Guidorzi, Hill, Kennea,
  Nousek, Osborne, \& Tagliaferri}]{moretti_new_2009}
Moretti, A., Pagani, C., Cusumano, G., {et~al.} 2009, A\&A, 493,
  \dodoi{10.1051/0004-6361:200811197}

\bibitem[{Oh {et~al.}(2018)Oh, Koss, Markwardt, Schawinski, Baumgartner,
  Barthelmy, Cenko, Gehrels, Mushotzky, Petulante, Ricci, Lien, \&
  Trakhtenbrot}]{oh_105-month_2018}
Oh, K., Koss, M., Markwardt, C.~B., {et~al.} 2018, ApJS, 235,
  \dodoi{10.3847/1538-4365/aaa7fd}

\bibitem[{Olausen \& Kaspi(2014)}]{olausen_mcgill_2014}
Olausen, S.~A., \& Kaspi, V.~M. 2014, ApJS, 212,
  \dodoi{10.1088/0067-0049/212/1/6}

\bibitem[{Schanne(2009)}]{schanne_presentation_2009}
Schanne, S. 2009, Presentation on the {{ECLAIRs Science}} Meeting ({{IAP}},
  2009/01/09)

\bibitem[{Schanne {et~al.}(2015)Schanne, Cordier, Atteia, Godet, Lachaud, \&
  Mercier}]{schanne_eclairs_2015}
Schanne, S., Cordier, B., Atteia, J.-L., {et~al.} 2015, in Proceedings of
  {{Swift}}: 10 {{Years}} of {{Discovery}} \textemdash{} {{PoS}}({{SWIFT}} 10),
  Vol. 233 ({SISSA Medialab}), \dodoi{10.22323/1.233.0107}

\bibitem[{Schanne {et~al.}(2013)Schanne, Le~Provost, Kestener, Gros, Cortial,
  G{\"o}tz, Sizun, Ch{\^a}teau, \& Cordier}]{schanne_scientific_2013}
Schanne, S., Le~Provost, H., Kestener, P., {et~al.} 2013, in 2013 {{IEEE
  Nuclear Science Symposium}} and {{Medical Imaging Conference}},
  \dodoi{10.1109/NSSMIC.2013.6829408}

\bibitem[{Schanne {et~al.}(2019)Schanne, Dagoneau, Ch{\^a}teau, Le~Provost,
  Daly, Anvar, Antier, Gros, \& Cordier}]{schanne_svom_2019}
Schanne, S., Dagoneau, N., Ch{\^a}teau, F., {et~al.} 2019, Mem. Soc. Astron.
  Ital., 90.
\newblock \url{http://adsabs.harvard.edu/abs/2019MmSAI..90..267S}

\bibitem[{Sizun(2011)}]{sizun_synthesis_2011}
Sizun, P. 2011, Synthesis of {{ECLAIRs Geant4}} Simulations, Tech. rep.

\bibitem[{Toor \& Seward(1974)}]{toor_crab_1974}
Toor, A., \& Seward, F.~D. 1974, AJ, 79, \dodoi{10.1086/111643}

\bibitem[{Wei {et~al.}(2016)Wei, Cordier, Antier, Antilogus, Atteia, Bajat,
  Basa, Beckmann, Bernardini, Boissier, Bouchet, Burwitz, Claret, Dai, Daigne,
  Deng, Dornic, Feng, Foglizzo, Gao, Gehrels, Godet, Goldwurm, Gonzalez,
  Gosset, G{\"o}tz, Gouiffes, Grise, Gros, Guilet, Han, Huang, Huang, Jouret,
  Klotz, La~Marle, Lachaud, Le~Floch, Lee, Leroy, Li, Li, Li, Liang, Lyu,
  Mercier, Migliori, Mochkovitch, O'Brien, Osborne, Paul, Perinati, Petitjean,
  Piron, Qiu, Rau, Rodriguez, Schanne, Tanvir, Vangioni, Vergani, Wang, Wang,
  Wang, Wang, Watson, Webb, Wei, Willingale, Wu, Wu, Xin, Xu, Yu, Yu, Yu,
  Zhang, Zhang, Zhang, \& Zhou}]{wei_deep_2016}
Wei, J., Cordier, B., Antier, S., {et~al.} 2016.
\newblock \url{https://arxiv.org/abs/1610.06892}

\bibitem[{Weisskopf {et~al.}(2010)Weisskopf, Guainazzi, Jahoda, Shaposhnikov,
  O'Dell, Zavlin, {Wilson-Hodge}, \& Elsner}]{weisskopf_calibrations_2010}
Weisskopf, M.~C., Guainazzi, M., Jahoda, K., {et~al.} 2010, ApJ, 713,
  \dodoi{10.1088/0004-637X/713/2/912}

\bibitem[{Wenger {et~al.}(2000)Wenger, Ochsenbein, Egret, Dubois, Bonnarel,
  Borde, Genova, Jasniewicz, Lalo{\"e}, Lesteven, \&
  Monier}]{wenger_simbad_2000}
Wenger, M., Ochsenbein, F., Egret, D., {et~al.} 2000, A\&AS, 143,
  \dodoi{10.1051/aas:2000332}

\bibitem[{{Wilson-Hodge} {et~al.}(2012){Wilson-Hodge}, Case, Cherry, Rodi,
  {Camero-Arranz}, Jenke, Chaplin, Beklen, Finger, Bhat, Briggs, Connaughton,
  Greiner, Kippen, Meegan, Paciesas, Preece, \& {von
  Kienlin}}]{wilson-hodge_three_2012}
{Wilson-Hodge}, C.~A., Case, G.~L., Cherry, M.~L., {et~al.} 2012, ApJS, 201,
  \dodoi{10.1088/0067-0049/201/2/33}

\bibitem[{Zhao {et~al.}(2012)Zhao, Cordier, Sizun, Wu, Dong, Schanne, Song, \&
  Liu}]{zhao_influence_2012}
Zhao, D., Cordier, B., Sizun, P., {et~al.} 2012, Exp Astron, 34,
  \dodoi{10.1007/s10686-012-9313-2}

\end{thebibliography}

\end{document}